\documentclass[acmsmall]{acmart}
\usepackage{graphicx}
\usepackage{subcaption}
\usepackage{caption}
\usepackage[export]{adjustbox} 
\usepackage{booktabs}
\usepackage{array}
\usepackage{ulem}
\usepackage{wrapfig}
\usepackage{enumitem}
\usepackage{paralist}
\usepackage{epstopdf}

\AtBeginDocument{%
  }

\renewcommand\footnotetextcopyrightpermission[1]{} 
\setcopyright{none} 
\usepackage{fancyhdr}
\makeatletter

\makeatletter
\renewcommand{\@journalNameShort}{Custom Journal Name}
\renewcommand{\@acmVolume}{Custom Volume}
\renewcommand{\@acmNumber}{Custom Issue Number}
\renewcommand{\@acmPubDate}{Custom Publication Date}
\makeatother

\makeatletter
\AtBeginDocument{%
  \fancypagestyle{standardpagestyle}{%
    \fancyfoot{} 
  }

  \fancypagestyle{firstpagestyle}{%
    \fancyhf{}
  }

  \thispagestyle{firstpagestyle} 
  \pagestyle{standardpagestyle} 
}
\makeatother

\begin{document}

\title{Orderly Management of Packets in RDMA by Eunomia}

\author{Sana Mahmood}
\email{smahmo12@jhu.edu}
\orcid{0000-0002-1486-3953}
\affiliation{%
  \institution{Johns Hopkins University}
  \city{Baltimore}
  \state{Maryland}
  \country{USA}
}

\author{Jinqi Lu}
\email{jlu114@jhu.edu}
\affiliation{%
  \institution{Johns Hopkins University}
  \city{Baltimore}
  \state{Maryland}
  \country{USA}
}

\author{Soudeh Ghorbani}
\email{sghorba1@jhu.edu}
\affiliation{%
  \institution{Johns Hopkins University}
  \city{Baltimore}
  \state{Maryland}
  \country{USA}
}


\renewcommand{\shortauthors}{Sana et al.}



\begin{abstract}
    To fulfill the low latency requirements of today’s applications, deployment of RDMA in datacenters has become prevalent over the recent years. However, the in-order delivery requirement of RDMAs prevents them from leveraging powerful techniques that help improve the performance of datacenters, ranging from fine-grained load balancers to throughput-optimal expander topologies. We demonstrate experimentally that these techniques significantly deteriorate the performance in an RDMA network because they induce packet reordering. Furthermore, lifting the in-order delivery constraint enhances the flexibility of RDMA networks and enables them to employ these performance-enhancing techniques. To realize this, we propose an ordering layer, Eunomia, to equip RDMA NICs to handle packet reordering. Eunomia employs a hybrid-dynamic bitmap structure that efficiently uses the limited on-chip memory with the help of a customized memory controller and handles high degrees of packet reordering. We evaluate the feasibility of Eunomia through an FPGA-based implementation and its performance through large-scale simulations. We show that Eunomia enables a wide range of applications in RDMA datacenter networks, such as fine-grained load balancers which improve performance by reducing average flow completion times by 85\% and 52\% compared to ECMP and Conweave, respectively, or employment of RDMA in expander topologies like Jellyfish which allows up to 60\% lower flow completion times and higher throughput gains compared to Fat tree. 
    
\end{abstract}

\ccsdesc[500]{Networks~Data center networks}

\keywords{RDMA, Load Balance, Packet Reordering}


\maketitle

\section{Introduction}
Over the recent years, RDMA over Ethernet (ROCEv2) has become prevalent in datacenter networks \cite{li2019hpcc, guo2016rdma}, owing to the significant reduction in host-side latency that it provides by bypassing the kernel and offloading operations on the RDMA Network Interface Cards (RNICs). RDMAs have strict requirements for losslessness and in-order delivery of data. Flow control techniques such as Priority flow control (PFC) \cite{pfc} and BFC \cite{bfc} strive to deliver the losslessness requirement and their impact on RDMAs' performance has been extensively studied \cite{li2019hpcc, guo2016rdma, bfc}. In contrast, the impact of the in-order delivery requirement is relatively unknown. In this work, we demonstrate that this requirement acts as a major roadblock for deploying the performance optimization techniques proven to enhance the efficiency of TCP/IP datacenters such as throughput-efficient expander graphs \cite{valadarsky2016xpander, singla2012jellyfish}, fine-grained load balancers (LBs) \cite{rps, ghorbani2017drill, presto, homa, ndp}, deflection for burst mitigation \cite{abdous2021burst, canary, vanini2017let, dibs}, sophisticated flow scheduling \cite{alizadeh2013pfabric} and fast failure recovery \cite{vanini2017let}. We propose a design that lifts this constraint and highlights the performance gains achieved through it.

The requirement for in-order delivery in RDMA arises from offloading CPU operations on the limited resource NIC that can not support sophisticated congestion control and loss recovery techniques such as TCP. When receiving an out-of-order (OOO) packet, the state-of-the-art congestion control algorithms for RDMAs such as HPCC \cite{li2019hpcc} and DCQCN \cite{zhu2015congestion} rely on simple and resource-efficient recovery mechanism such as Selective Repeat \cite{mittal2018revisiting} (SR) and go-back-N (GBN). Both recovery mechanisms consider an out-of-order packet to indicate a packet loss and move the send window to the first unacknowledged packet. GBN then retransmits all packets while SR retransmits only the unacknowledged packets. Such recovery mechanisms are suitable for networks with little to no packet reordering, and fail to keep up if the network induces a high degree of reordering. This renders the existing RDMA network inflexible towards adopting existing or newer techniques that induce reordering but can provide significant performance gains.

A plethora of research on datacenter networking takes advantage of the relative resilience of TCP/IP networks towards limited degrees of packet reordering to optimize performance \cite{rps, ghorbani2017drill, alizadeh2014conga, abdous2021burst, canary, vanini2017let, dibs, alizadeh2013pfabric, pabo, presto, homa}. These techniques leverage the transport support for reordered packets (e.g., TCP reorder buffer) or offload the packet reordering support to a below-transport software layer in the host stack \cite{abdous2021burst, ghorbani2017drill, presto}. For example, 1) many fine-grained LBs \cite{ghorbani2017drill, alizadeh2014conga, vanini2017let} that allow the network to leverage the multi-path datacenter topologies to improve performance and manage failures, 2) expander topologies with lower average path lengths that deliver better performance and easier extension \cite{singla2012jellyfish, valadarsky2016xpander}, 3) packet deflection which defuses microbursts and mitigates the hotspots in the network \cite{abdous2021burst, dibs, pabo, canary}, and 4) flow scheduling for aiding latency-sensitive flows \cite{alizadeh2013pfabric} all have proven to deliver high-performance gains in traditional TCP/IP datacenter networks but induce a high degree of packet reordering. For example, in fine-grained LBs, packets on slower paths will arrive out-of-order, similarly in expander topologies, different paths have different lengths and since these topologies rely on sub-flow level or finer-grained routing, packets are bound to arrive out-of-order. Lastly, deflection diverts packets from hotspots to take longer routes and arrive out-of-order. These techniques leverage \emph{software} (e.g., below the transport layer \cite{abdous2021burst}) to manage reordering. In RDMAs, however, transport protocols are hardware-resident and fundamentally more constrained, making it hard to handle packet reordering. 

Packet reordering makes employing these techniques in RDMA datacenter networks challenging. We show, for example, that while fine-grained load balancing techniques such as DRILL \cite{ghorbani2017drill} improve the flow completion times (FCT) in TCP/IP datacenters by $1.6\times$ compared to ECMP, in an RDMA network, their performance \emph{is degraded} by \emph{$9.5\times$} in comparison with ECMP. This is because RDMA inherently can not handle a high degree of packet reordering, and the mechanisms that are used in TCP/IP networks to mitigate the impact of packet reordering, such as ordering packets in the below-transport layer host stack \cite{abdous2021burst} can not be used in RDMA. RDMA implements transport on the NIC, bypasses the kernel, and places data directly in application memory, which leaves no room for ordering packets (except for on the NIC) without triggering its recovery mechanism.

We argue that ordering support in RDMA will be a key enabler for leveraging much of the vast literature on datacenter performance enhancement. We show through experiments that lifting the in-order delivery constraint makes RDMA datacenters flexible and enables them to benefit from techniques such as fine-grained LBs, expanders, deflection, flow scheduling, and real-time failover methods. To realize this vision, we present an ordering layer for RDMA NICs called Eunomia.\footnote{Greek Goddess of order} Eunomia is implemented on the RDMA NICs and tracks the packet ordering of all incoming flows. Eunomia uses a hybrid-dynamic (HD) bitmap data structure to track the order of packets, each packet/sequence number corresponds to a bit in the bitmap. The bitmap is organized in a hybrid fashion of circular and linear arrays, to avoid wasteful utilization of limited on-chip memory, and its size is dynamic. We develop a memory controller for the NIC to provide dynamic allocation and access of memory for HD bitmap. 
Furthermore, we evaluate the feasibility of our design through an FPGA implementation using the AMD Kintex UltraScale+ FPGA KCU116 base model. Our evaluation includes the simulation of three key modules: the packet driver, memory controller, and HD bitmap. The results demonstrate a significant reduction in average bitmap memory utilization compared to a naïve static bitmap implementation. This demonstrates Eunomia’s suitability for resource-constrained environments and offers substantial advantages over existing static bitmap-based solutions such as IRN \cite{mittal2018revisiting} and MPRDMA \cite{lu2018multi}. 

Equipping RDMA datacenters with ordering support is a doorway toward leveraging vast literature on datacenter networks to improve performance. Eunomia acts as that doorway and enables RDMA datacenter networks to deploy techniques such as fine-grained LBs, deflection, flow scheduling, and leverage irregular graph-based topologies, which help improve performance. For example, our large-scale simulation experiments demonstrate that in a 1:2 oversubscribed Clos topology with 8 spine and 8 leaf switches, at a high network load (80\%) Eunomia reduces the mean FCT up to 85\%, and 52\% compared to ECMP and Conweave \cite{song2023network}, respectively. Eunomia also allows RDMA to reduce the PFC pause duration on ports by up to $\approx$99\% compared to the state-of-the-art.

\section{Inflexibility of RDMA: Lack of Ordering Support}
RDMA datacenter networks provide significant performance benefits compared to traditional TCP/IP networks, reportedly achieving 7.8 times lower latencies for latency-sensitive services \cite{mittal2018revisiting}. However, they are inflexible in borrowing techniques that are known to improve performance in traditional TCP/IP networks, largely due to the in-order delivery constraint. We demonstrate through extensive simulation experiments that this constraint limits RDMAs' ability to employ techniques such as fine-grained LBs, which not only improve performance but also enable further applications like irregular graph topologies and failure management. We also demonstrate that lifting this constraint makes RDMA flexible and allows for considerable performance gains through leveraging these techniques. It is worth noticing that the advantages of lifting in-order delivery constraint are not limited to the techniques mentioned below, illustrated in \S \ref{sec:eval}.

\begin{figure}[t]
\begin{minipage}[b]{0.69\linewidth}
  \centering
  \begin{subfigure}{\linewidth}
    \centering
    \includegraphics[width=\linewidth]{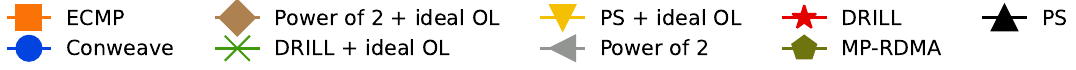}
  \end{subfigure}

  \begin{subfigure}{0.32\linewidth}
    \centering
    \includegraphics[width=\linewidth]{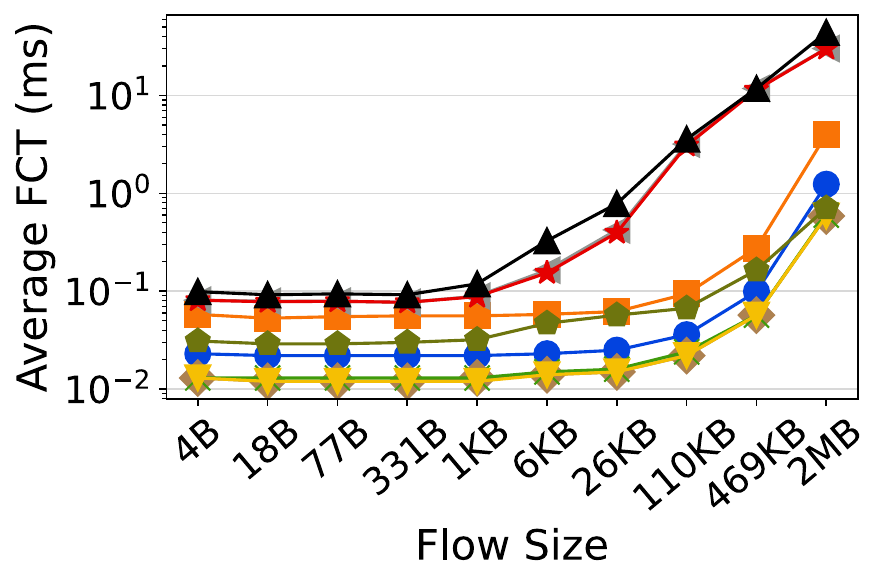}
    \caption{Lossless Network}
    \label{fig:motivation_clos_avg_fct_80p}
  \end{subfigure}
  \hfill
  \begin{subfigure}{0.32\linewidth}
    \centering
    \includegraphics[width=\linewidth]{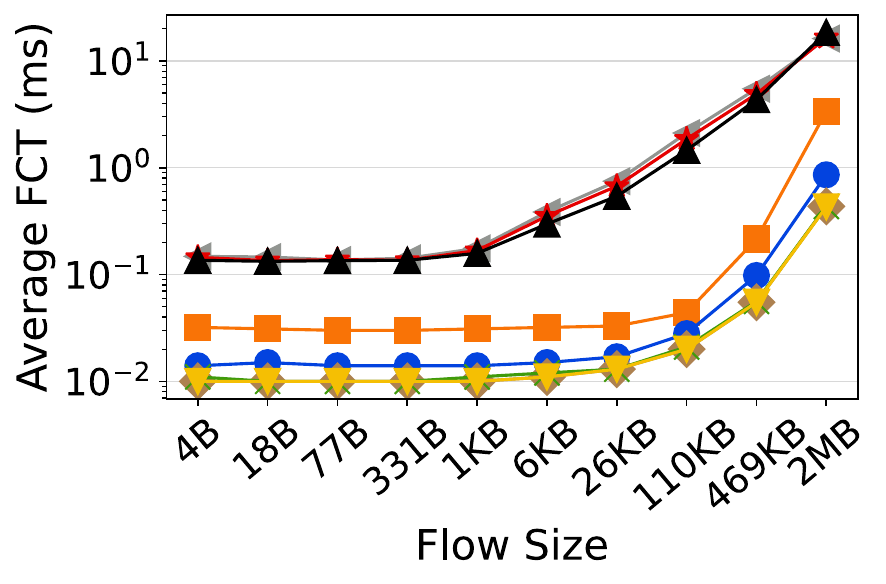}
    \caption{Lossy Network (IRN)}
    \label{fig:motivation_clos_avg_fct_80p_irn}
  \end{subfigure}
  \hfill
  \begin{subfigure}{0.32\linewidth}
    \centering
    \includegraphics[width=\linewidth]{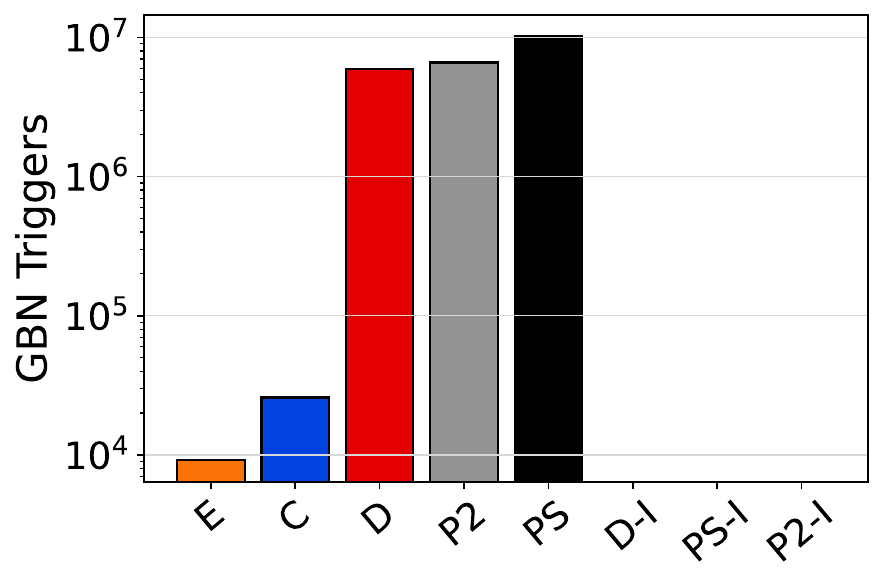}
    \caption{Lossless Network}
    \label{fig:motivation_clos_gbn_80p}
  \end{subfigure}
  \vspace{-2mm}
  \caption{Fine-grained LBs significantly deteriorate in performance in (a) lossless RDMA and (b) lossy RDMA due to (c) heavy reordering, and pairing them with an ordering layer allows them to reach their full potential.}
  \label{fig:motivation_clos}
  \end{minipage}
  \hfill
    \begin{minipage}[b]{0.29\linewidth}
        \centering
        \begin{subfigure}{\linewidth}
            \centering
            \includegraphics[width=\linewidth]{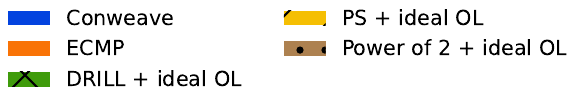}
        \end{subfigure}
        \begin{subfigure}[b]{\linewidth}
            \centering
            \includegraphics[width=0.7\linewidth]{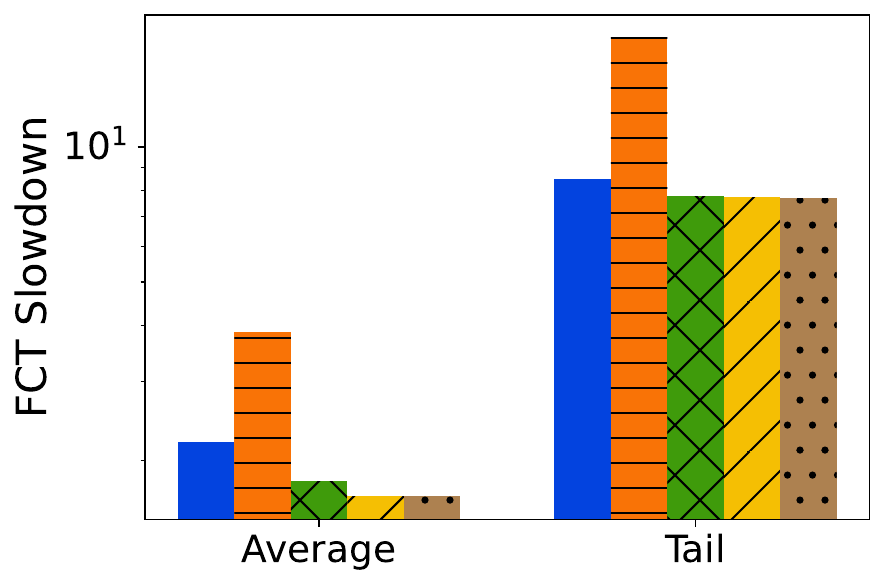}
            \label{fig:motivation_failure_clos_tail_50p}
        \end{subfigure}
        \vspace{-5mm}
        \caption{Ordering support enables RDMA networks to maintain their performance under link failures in a Clos topology.}
        \label{fig:motivation_failure}
    \end{minipage}
  \vspace{-3mm}
\end{figure}

\subsection{Leveraging multi-paths in datacenter topologies}
\label{sec:motivation_lb_comp}

Most commonly deployed datacenter topologies such as Clos or Fat tree topology \cite{180325} offer path redundancy in the form of equal cost multiple paths. Over the years, different types of LBs have been proposed for traditional datacenter networks to leverage those multi-path topologies. Most of these LBs work at finer grain such as flowlet, sub-flow, or packet level \cite{ghorbani2017drill, alizadeh2014conga, vanini2017let, presto, rps}, and consequently induce some degree of packet reordering in the network. Since RDMA is not inherently equipped to handle packet reordering, it can not employ such fine-grained LBs and relies on ECMP (a per-flow hash-based routing), which fails to leverage the multi-path network to its fullest due to known underlying issues, such as hash collisions \cite{vanini2017let}. We validated this hypothesis via simulation experiments in ns3 \cite{ns3_simul}, on an 8x8 Clos topology with 128 servers (1:2 over-subscription), 100G links, and 8us round trip time (RTT). We used AliStorage \cite{li2019hpcc} distribution to generate traffic. We employed a few fine-grained LBs such as DRILL \cite{ghorbani2017drill}, Power of 2 (which picks 2 random paths and sends the packet to least loaded) \cite{ghorbani2017drill} and packet spray (PS, a round-robin version of \cite{rps}), and compared them against ECMP, Conweave \cite{song2023network} (a sub-flow level LB for RDMA that rearranges OOO packets in the network) and, MP-RDMA \cite{lu2018multi} (a multi-path transport for RDMA similar to \cite{raiciu2011improving}). We tested this in both a lossless (with GBN) and a lossy (with IRN\cite{mittal2018revisiting}) network. We see that all fine-grained LBs perform significantly worse than existing techniques (Fig \ref{fig:motivation_clos}). For example, DRILL \cite{ghorbani2017drill} results in $9.5\times$ higher mean FCT than ECMP, whereas in a TCP/IP network it lowers FCT by $1.6\times$ than ECMP \cite{ghorbani2017drill}. This behavior can be attributed to the lack of packet ordering support in RDMA NICs. An out-of-order packet is believed to occur due to packet loss and the go-back-n (GBN) (or selective repeat in Fig \ref{fig:motivation_clos_avg_fct_80p_irn}) mechanism is triggered for recovery, we see this in Fig \ref{fig:motivation_clos_gbn_80p} (first letter of the LB is used as an acronym in the graph due to limited space, "I" indicates Ideal ordering layer), where a large number of GBN triggers occur in finer-grained LBs without ordering support.

To validate that packet reordering is indeed the culprit in degrading performance. We equipped the RDMA network with an ideal ordering layer (ideal OL) which sits underneath the NIC, buffers the out-of-order packets, and delivers them only after putting them in order. When equipped with an ideal ordering layer, the performance of finer-grained LBs improves significantly compared to existing techniques such as ECMP, Conweave \cite{song2023network} and MP-RDMA \cite{lu2018multi} both in lossless and lossy RDMA. In Fig \ref{fig:motivation_clos_avg_fct_80p}, DRILL with an ideal ordering layer performs $6.1\times$ better than ECMP, because fine-grained LBs spread the load across available paths effectively, consequently improving performance when packet reordering is taken care of.

\subsection{Employing RDMA in expander topologies} 
Compared to common multi-rooted datacenter topologies (e.g., Clos and fat-tree), expander graphs deliver higher throughput and lower latency thanks to their lower average path lengths \cite{singla2012jellyfish, valadarsky2016xpander}. Can RDMAs, too, take advantage of their higher performance? We show that in an RDMA deployment with no ordering support, the answer to this question is negation. 

\begin{figure}[t]
    \centering
    \begin{minipage}[b]{0.49\linewidth}
        \centering
        \includegraphics[width=0.8\linewidth]{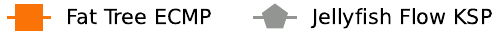}
        
        \begin{subfigure}[b]{0.49\linewidth}
            \centering
            \includegraphics[width=0.9\linewidth]{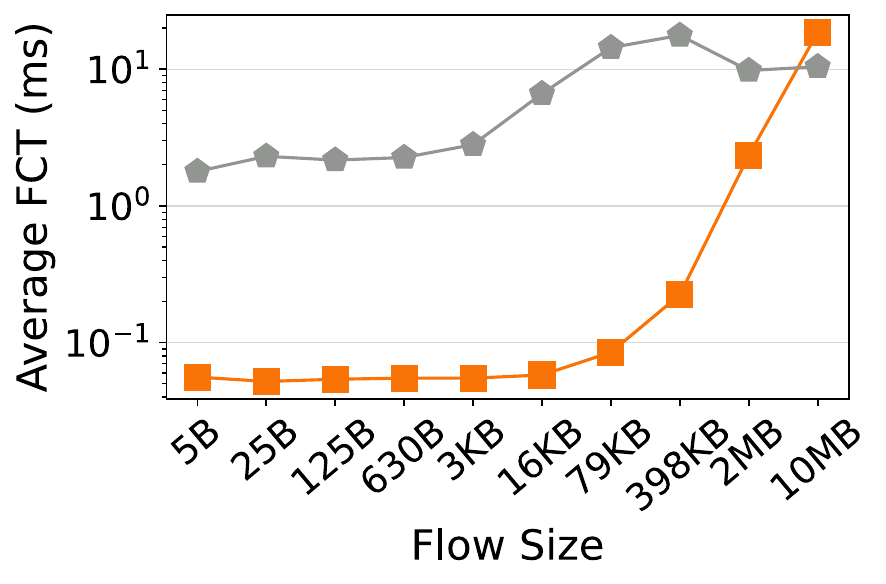}
            \caption{Avg FCT (80\% Load)}
            \label{fig:avg_fct_jellyfish_80p_motivation}
        \end{subfigure}
        \hfill
        \begin{subfigure}[b]{0.49\linewidth}
            \centering
            \includegraphics[width=0.9\linewidth]{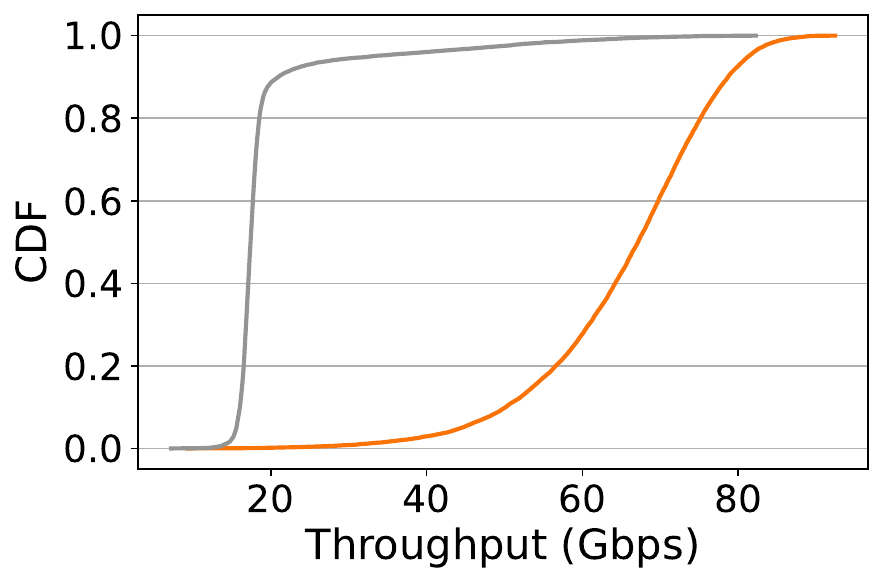}
            \caption{Throughput (80\% Load)}
            \label{fig:thrpt_jellyfish_80p_motivation}
        \end{subfigure}
        \caption{Employing RDMA without ordering support in Jellyfish topology significantly degrades performance.}
        \label{fig:motivation_jellyfish}
    \end{minipage}
    \hfill
    \begin{minipage}[b]{0.49\linewidth}
        \centering
        \begin{subfigure}[b]{1.0\linewidth}
            \centering
            \includegraphics[width=0.8\linewidth]{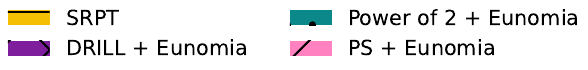}
            \label{fig:ooo_legend}
        \end{subfigure}
        \begin{subfigure}[b]{0.49\linewidth}
            \centering
            \includegraphics[width=0.9\linewidth]{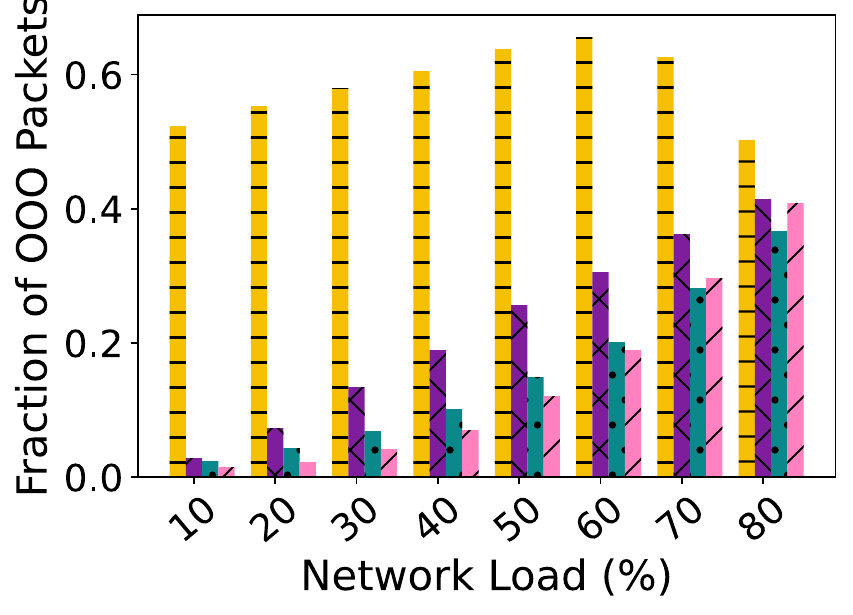}
            \caption{Load vs Reordering}
            \label{fig:ooo_fraction_vs_load}
        \end{subfigure}
        \hfill
        \begin{subfigure}[b]{0.49\linewidth}
            \centering
            \includegraphics[width=0.9\linewidth]{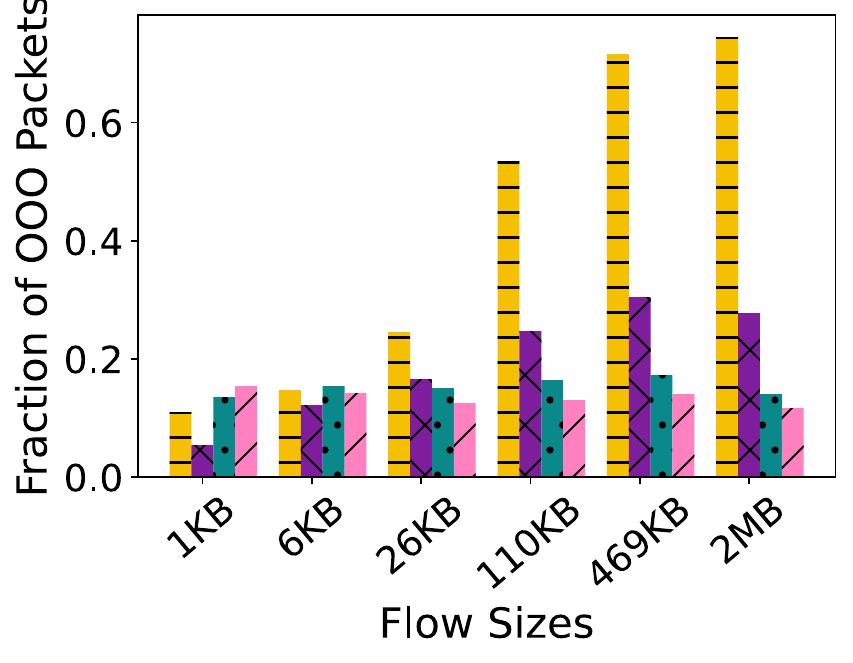}
            \caption{Size vs Reordering}
            \label{fig:ooo_fraction_vs_fsize}
        \end{subfigure}
        \caption{Network factors impacting packet reordering}
        \label{fig:ooo_fraction_comparison}
    \end{minipage}
    \vspace{-2mm}
\end{figure}

Expander graph topologies rely on sub-flow level or fine-grained routing techniques to fully leverage their rich path diversity. They fail to deliver promised performance with a flow level routing \cite{singla2012jellyfish, valadarsky2016xpander}. We test this in ns-3 simulations and compare an 8-ary Fat tree topology, with an equivalent Jellyfish topology \cite{singla2012jellyfish} of 80 switches and 128 servers. All links are 100G with 1us propagation delay. In the Fat tree, the baseline is ECMP, and for Jellyfish we used flow level k-shortest path (FKSP) routing, which sends the whole flow on a randomly selected path. FKSP gives us a baseline of directly deploying RDMA in Jellyfish (without the need for ordering support). In Fig \ref{fig:motivation_jellyfish}. We see that the performance of Jellyfish + FKSP is significantly worse than Fat tree because a flow-level routing is not suitable for variable path lengths. Furthermore, in Fig \ref{fig:thrpt_jellyfish_80p_motivation} we see that Jellyfish achieves much lower throughput than Fat tree because a flow-level routing (FKSP) fails to efficiently utilize the available paths, highlighting the need for fine-grained routing in Jellyfish. These results indicate that RDMA can not be directly deployed in expander graph topologies.

\subsection{Managing Failures}
Link failures are frequent in datacenters and can cause significant performance loss \cite{gill2011understanding, 180325}. To manage failures, different failover techniques are employed such as path redundancy in topologies \cite{180325}, fine-grained load balancing, and recovery mechanism \cite{ghorbani2017drill, alizadeh2014conga, raiciu2011improving}. Efficient failure management in datacenter switches can induce packet reordering in a flow. Since RDMA inherently can not handle packet reordering, it can exacerbate the performance loss on flows affected by link failures.

We evaluated the impact of link failures on performance (same setup as \S \ref{sec:motivation_lb_comp}) and compared different LBs under failure. We repeatedly failed 1\% of the network links at fixed intervals of time. After each interval the failed links come back up and new links fail. Upon a link failure, the switches drain the buffers of failed link ports and recalculate the routing tables. We only observed the flows that are affected by the failures, that is the ones that face at least one failure event. Fig \ref{fig:motivation_failure} shows that coarse-grained LBs are affected more and under-perform compared to the fine-grained LBs combined with an ideal ordering layer. In fine-grained LBs packets of a flow take different paths reducing the probability of multiple packets facing failure. This allows fine-grained LBs to maintain their performance in the face of link failures and degrade gracefully (Fig \ref{fig:motivation_failure}). Coarse-grained LBs, on the other hand, run at risk of losing a large chunk of packets if a flow's path experiences link failure. This would require multiple re-transmissions to recover the lost packets and result in degraded performance, as can be seen in Fig \ref{fig:motivation_failure}. Therefore, ordering support in RDMA can open avenues for incorporating effective and efficient failure management techniques.

In summary, we demonstrated that the lack of ordering support in RDMA networks hinders them from achieving high performance compared to the state-of-the-art techniques. An RDMA network equipped with ordering support has the potential to achieve 1) $\approx81\%$ better performance in terms of flow completion times by employing finer-grained LBs 2) reap the benefits of expander graph topologies, and 3) employ better failure management. Furthermore, the advantages spread beyond the applications enabled by fine-grained LBs as we demonstrate in \S\ref{sec:eval}.

\section{Design}
In order to equip RDMA network to handle out-of-order (OOO) packets, we present Eunomia, an ordering layer that resides on the NIC, and ensures an in-order delivery of data to the application.

Handling reordering in RDMA is challenging because RDMA NICs are known to have limited on-chip memory. For example, while SRAM size of commercial RNICs is proprietary, some works \cite{wang2023srnic, kalia2019datacenter} have reported that common chips such as Mellanox CX-5 \cite{nvidia-connectx5} have an on-chip memory of $\approx$2MB. Thus, a reordering design needs to use memory efficiently. A straightforward approach would be not to store complete packets on the NIC, instead place them in the host memory and use a pointer-like bitmap data structure to serve as a reordering buffer for each connection on the NIC \cite{mittal2018revisiting}\cite{lu2018multi}. This approach comes with a trade-off between memory taken on the NIC vs the degree of reordering handled per connection. A larger reordering bitmap per connection requires more memory but also handles high reordering. Furthermore, calculating the optimal size for the reordering bitmap is non-trivial as we demonstrate through experiments in \S\ref{sec:dynamic_reordering_bitmap}. To resolve these challenges, Eunomia is built on the intuition that the reordering bitmap for a connection should only be allocated as the need arises and should dynamically increase in size if a connection faces high reordering. To realize this, Eunomia introduces two key modules, a \textbf{Hybrid-Dynamic Bitmap} structure which dynamically grows in size as the degree of reordering increases for a connection, and a \textbf{Memory Controller}, which allows dynamic allocation of memory to the bitmap module on the fly. These modules allow Eunomia to preserve memory when connections do not face reordering and also efficiently allocate memory in the face of reordering.

\subsection{Overview}
Eunomia sits on the NIC where it intercepts all data packets of a connection, tracks the state of packet reordering for each connection, and generates an appropriate response, before placing them in the host memory. Eunomia begins with a \textbf{\textit{Sender-Side Agent}} which processes the outgoing data packets by appending additional metadata to them, the metadata includes the first sequence number of the connection and a bit to indicate whether it is the last packet.
At the receiver, a \textbf{\textit{Receiver-Side Agent}} intercepts the incoming data packets and sends them to the \textbf{\textit{Hybrid-Dynamic Bitmap Module}}, where the OOO packets are tracked before getting placed in the application memory. Hybrid-Dynamic bitmap module interacts with a \textbf{\textit{Memory Controller}} to dynamically allocate bitmaps for a connection. The receiver-side agent then generates an ACK/SACK/NACK based on the packets' order and whether it was successfully tracked in the dynamic bitmap. At the sender, the agent intercepts the incoming ACK/SACK/NACK and ensures the acknowledgments of OOO packets do not trigger the loss recovery. Note that Eunomia's acknowledgments are not related to and interact with congestion control. Eunomia can interoperate with different congestion control algorithms (including DCQCN, a congestion control that does not rely on acknowledgments \S\ref{sec:eval}). Furthermore, these acknowledgments create a small bandwidth overhead ($\approx5\%$ of the total bandwidth from experiments in \S \ref{sec:eval_fine_grain_lb}). 
The rest of this section explains each module of Eunomia.

\subsection{Sender-Side Agent}
\label{sec:sender_side}

On the sender side, Eunomia only requires a couple of operations performed, one at the time of sending the data packet and another upon reception of the control packet (ACK/SACK/NACK).

\subsubsection{Appending Metadata}
When sending a data packet, Eunomia appends some extra metadata to each packet which is necessary for correctly handling packet reordering at the receiver. The metadata is appended by extending the RDMA header by 33 bits, 32 bits are used to append the first sequence number of the connection to each data packet, also referred to as \textit{Head}. The sequence number of the packet itself is already contained in the RDMA header, so Eunomia does not need to append that. The remaining 1 bit is used to indicate whether it is the last packet of the connection. 

\subsubsection{Reaction to Acknowledgment}
Upon receiving acknowledgments, Eunomia needs to ensure that recovery (GBN or selective repeat for example) does not get triggered by OOO packets' acknowledgments. Packets received OOO at the receiver are acknowledged using SACK, which contains the expected in-order sequence number and the sequence number of the packet received OOO. Upon receiving SACK, Eunomia does not trigger the recovery. Eunomia marks the SACKed sequence number as received (if needed), for example in selective repeat recovery offered by existing NICs the SACKed packets are marked received at the sender, and in GBN they are ignored. Upon receiving a NACK, which indicates that the packet could not be handled at the receiver, Eunomia triggers the underlying recovery mechanism.

\subsubsection{Loss Recovery}
RDMA assumes a congestion-related lossless network, but erroneous packet loss, albeit rare, can still occur. This can halt a naive implementation of a dynamic bitmap which assumes a missing (in this case lost) packet is merely out-of-order and will eventually be received. It will indefinitely wait for this packet or until the bitmap cap is reached, consequently degrading performance. 
We avoid this by leveraging the existing timeout mechanism at the sender NICs such as in \cite{mittal2018revisiting}, which triggers recovery (GBN or selective repeat) if nothing is heard from the destination host for the timeout duration.

\subsection{Hybrid-Dynamic Reordering Bitmap}
\label{sec:dynamic_reordering_bitmap}
We track OOO packets in Eunomia by using a hybrid-dynamic bitmap at the receiver NIC. In contrast to prior works that deploy static (fixed size) circular reordering bitmaps (RBs) \cite{lu2018multi}\cite{mittal2018revisiting}, we introduce the \emph{hybrid-dynamic (HD) bitmap}, which is a dynamically increasing RB (via the memory controller, discussed in \S\ref{sec:memory_controller}) and is managed as a combination of circular and linear arrays to ensure efficient and correct handling of OOO packets.

\subsubsection{Bitmap Size: } The size of the RB is a crucial parameter and can directly impact the performance. Keeping the limited-NIC memory in mind, existing works that use RB tend to have a limited fixed-size bitmap for all connections, and consequently handle a limited degree of reordering. Therefore, it is important to determine the appropriate size for the RB that handles reordering well and does not leave a large memory footprint. However, it is non-trivial to decide an optimal RB size that efficiently handles reordering because the degree of reordering is impacted by various factors that we discuss below. The experiments in Fig \ref{fig:ooo_fraction_comparison} use the same setup as \S\ref{sec:motivation_lb_comp} and demonstrate that it is non-trivial to calculate the optimal bitmap size at the time of its creation. 

\begin{compactitem}
    \item Network Load: Fig \ref{fig:ooo_fraction_vs_load} shows that the fraction of packets reordered increases with the network load, indicating that an optimal bitmap size changes as the network load varies.
    \item Flow Size: Fig \ref{fig:ooo_fraction_vs_fsize} shows the degree of reordering faced by different flow sizes. The larger the flow, the higher the reordering (2M is slightly lower because fewer flows are finished and accounted for). This result might suggest that the optimal RB size should be equal to the size of a flow. However, it imposes a condition that flow sizes are known apriori and will also require significantly large memory for larger flows.
    \item Network Settings: Different types of network settings, such as fine-grained LBs, flow scheduling and deflection, induce different degrees of packet reordering. For example, DRILL induces more reordering than other LBs, or flow scheduling like SRPT adds significantly high reordering, as evident in \ref{fig:ooo_fraction_comparison}. End hosts are usually agnostic to the network settings and therefore can not determine an optimal RB size that suits the network's needs.
\end{compactitem}

\subsubsection{Dynamic Size:} We equip Eunomia with a dynamically increasing RB, which starts with a small fixed size and is extended (by the same fixed size) as needed. Starting with a small size ensures minimal resource waste for flows that would not face high reordering and increasing the size dynamically ensures efficient handling of high reordering. However, Eunomia limits the maximum size that a dynamic bitmap can reach, to ensure no starvation (a few connections take up all of the space while the rest of them starve).

\subsubsection{Hybrid Management:} 
\begin{figure}[t]
        \centering
        \begin{subfigure}[b]{0.49\linewidth}
            \centering
            \includegraphics[width=\linewidth]{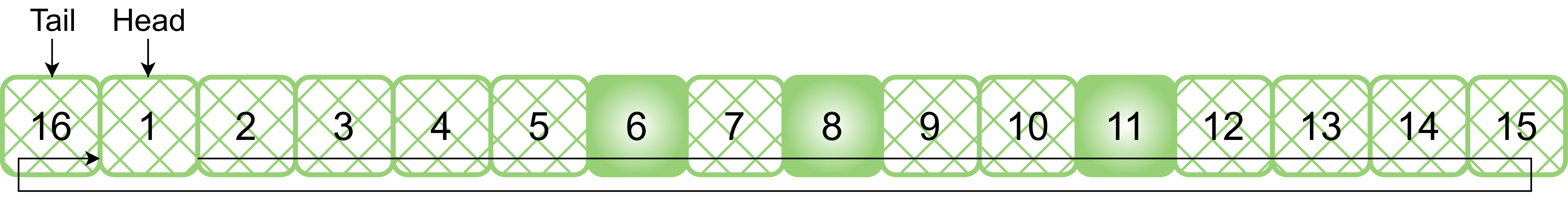}
            \caption{Received packet 0, 6, 8 and 11 in a circular bitmap}
            \label{fig:linear_hd_bm_1}
       \end{subfigure}
        \centering
        \hfill
        \begin{subfigure}[b]{0.49\linewidth}
        \centering
        \includegraphics[width=\linewidth]{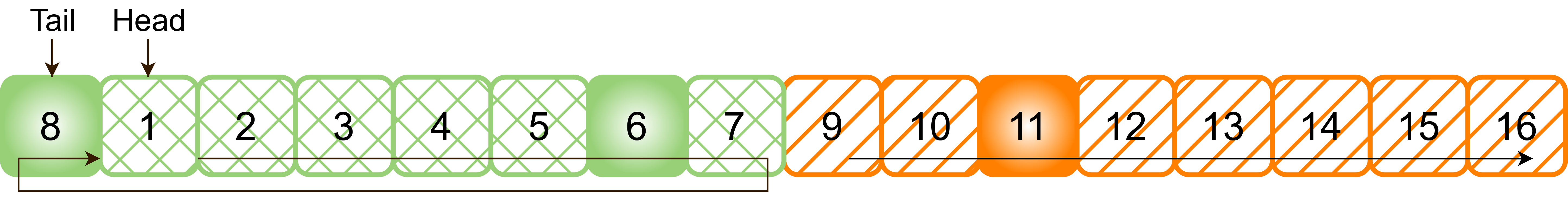}
        \caption{Received packet 0, 6, 8 and 11 in a hybrid bitmap}
        \label{fig:linear_hd_bm_2}
   \end{subfigure}
   \vspace{-2mm}
\label{fig:dynamic_bitmap_example_1D}
\caption{Hybrid-Dynamic Reordering Bitmap (Linear Representation) - Each block represents a bit in the bitmap array and the number inside represents the corresponding sequence number (consecutive for convenience). Single hatched is circular bitmap while double hatched is linear. If highlighted the bit is ON, otherwise OFF.}
\end{figure}

HD bitmap is organized primarily as a circular bitmap array, with a moving head corresponding to the first missing sequence number, and the mapping for subsequent sequence numbers is calculated using this head as a reference, shown in Fig \ref{fig:dynamic_table_subfigure1}. However, increasing the size of the circular bitmap changes the mapping of sequence numbers to bits, as shown in Fig \ref{fig:linear_hd_bm_1} which represents the bitmap in Fig \ref{fig:dynamic_table_subfigure1} after size increment upon receiving packet 11. This increment requires an expensive operation of looping through the whole bitmap array and updating values based on the new mapping. We avoid this by temporarily managing a part of the bitmap as a linear array whenever we increment size, as shown in Fig \ref{fig:linear_hd_bm_2}. The linear array is eventually absorbed into the circular array once the circular array is completely flushed. This hybrid approach allows Eunomia to utilize resources intelligently via circular management while avoiding remapping sequences upon size increment via linear management.

\subsubsection{Dynamic-Hybrid Bitmap Management}

\begin{figure}[t]
\begin{minipage}[h]{0.7\linewidth}
\small
\captionsetup{type=table} 
\caption{States kept for dynamic bitmap}
\label{tab:states}
\vspace{-2mm}
\begin{tabular}{|p{0.25\linewidth}|p{0.1\linewidth}|p{0.5\linewidth}|}
    \hline
    \textbf{Name} & \textbf{Size} & \textbf{Description} \\
    \hline
    \texttt{HD Bitmap Addr} & 4B & Address pointing to the HD bitmap \\
    \hline
    \texttt{Head} & 4B & First missing sequence number \\
    \hline
    \texttt{Tail} & 4B & Highest sequence number that circular bitmap can accommodate \\
    \hline
    \texttt{Last Seq} & 4B & Sequence number of last packet \\
    \hline
    \texttt{Head BM ID} & 1B & Bitmap ID that contains the head \\
    \hline
    \texttt{Head BM Index} & 1B & Index on the bitmap that contains head \\
    \hline
    \texttt{Circular BM Size} & 1B & Current size of the circular bitmap \\
    \hline
    \texttt{Dynamic Size} & 1B & Current total size of the bitmap \\
    \hline
    \texttt{Bitmap Blocks Relative Addr} & Relative & Relative address of each bitmap block allocated to the connection (dependent on block size and cap on bitmap size) \\
    \hline
  \end{tabular}
\end{minipage}
    \begin{minipage}[h]{0.29\linewidth}
        \centering
        \setcounter{subfigure}{0}
        \setlength{\belowcaptionskip}{0pt}
        \captionsetup{type=figure}
        \begin{subfigure}[t]{\linewidth}
            \centering
            \includegraphics[width=0.8\linewidth]{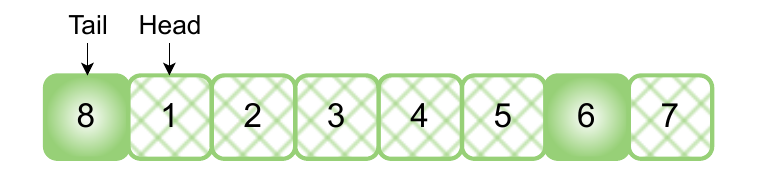}
            \caption{Received seq 0, 6, 8}
            \label{fig:dynamic_table_subfigure1}
        \end{subfigure}
        \begin{subfigure}[t]{\linewidth}
            \centering
            \includegraphics[width=0.8\linewidth]{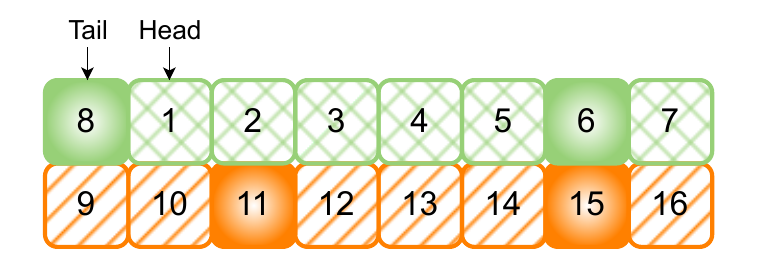}
            \caption{Received seq 11, 15}
            \label{fig:dynamic_table_subfigure2}
        \end{subfigure}
        \begin{subfigure}[t]{\linewidth}
            \centering
            \includegraphics[width=0.8\linewidth]{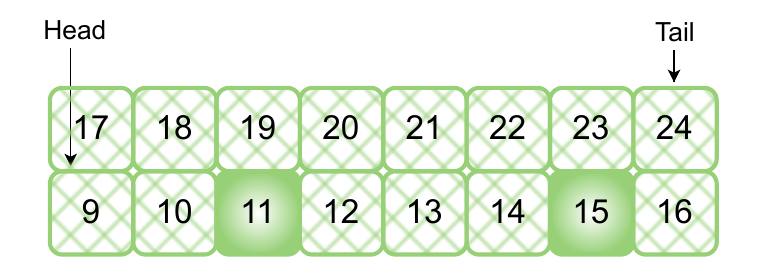}
            \caption{Received seq 1, 2, 3, 4, 5, 7}
            \label{fig:dynamic_table_subfigure3}
        \end{subfigure}
    \vspace{-4mm}
    \caption{Dynamic Bitmap Table}
    \label{fig:dynamic_table_example}
    \end{minipage}
\vspace{-6mm}
\end{figure}

We manage the HD bitmap using a logical table consisting of multiple fixed-sized small bitmaps as rows, shown in Fig \ref{fig:dynamic_table_subfigure3}. The size is increased by dynamically adding row(s) to the table. In such a table abstraction, a bit is referenced by the bitmap number (row) and the index on that bitmap (column). Either all or a chunk of rows from the top is part of the \textit{Circular Bitmap Portion}, which is managed as a circular array using $Head$ as the reference to calculate a bit corresponding to a sequence number. The bottom rows are part of the \textit{Linear Bitmap Portion} (if it exists), managed as a linear array where the sequence number following the $Tail$ corresponds to its first bit. To correctly map a sequence number to a bit in HD bitmap, we maintain few states at the receiver for each connection that faces reordering, mentioned in Table \ref{tab:states}.

Upon receiving a data packet, the receiver NIC checks whether it is the first OOO packet of a connection by using \textit{expected sequence} (a state maintained by receiver QP for transport) and \textit{HD Bitmap Addr}. If so, it creates an HD bitmap for the connection which consists of a single row of circular bitmap of \textit{Block Size}, a fixed bitmap block size. \textit{Head} points to the first expected sequence number in order, which is either \textit{Head}, carried in the packet as metadata, if the first packet of the connection is received OOO, or the \textit{expected sequence} number maintained by transport otherwise. Creating a bitmap only upon receiving an OOO packet ensures Eunomia does not allocate any bitmap memory (or metadata memory) for flows that face no reordering.

Now, if the sequence number of the packet arrived can be accommodated in the circular bitmap, its corresponding bit index is calculated using $Head$ and marked, as shown in Fig \ref{fig:dynamic_table_subfigure1}. If a packet arrives with a sequence number that can not be accommodated inside the bitmap, the size of the bitmap is increased by adding a linear bitmap of \textit{Block Size}. The \textit{Tail} continues to point to the highest sequence number that can be accommodated in the circular bitmap, and the first index of linear bitmap corresponds to the sequence number following the $Tail$, as shown in Fig \ref{fig:dynamic_table_subfigure2}. Once the size of the bitmap is adjusted, the index on the linear bitmap that corresponds to the sequence number is calculated using \textit{Tail} as a reference, and that bit is turned on (Fig \ref{fig:dynamic_table_subfigure2}). Furthermore, if a packet arrives with a sequence number that can not be accommodated in the circular bitmap, but there exists a linear bitmap that can absorb this sequence number, we perform the same steps for it as stated above (except for increasing the bitmap size), for example, $seq=15$ in Fig \ref{fig:dynamic_table_subfigure2}. Once the sequence number is traced in the bitmap, Eunomia checks whether the sequence number is equal to the $Head$ because this would indicate in-order delivery. If so, Eunomia resets the consecutive bitmap entries with value $=1$, starting from \textit{Head}. The \textit{Head} is then set to the first missing \textit{Seq}, as shown in Fig \ref{fig:dynamic_table_subfigure3}. Moreover, Eunomia checks whether all of the values in the circular bitmap portion are flushed (reset) and whether the bitmap contains a linear portion, if so, it merges the linear portion into the circular portion by updating \textit{Tail} and the \textit{Circular BM size} (Fig \ref{fig:dynamic_table_subfigure3}).

\subsection{Memory Controller}
\label{sec:memory_controller}

To implement an HD bitmap, we need to allocate memory on the fly, which is non-trivial in hardware. It requires a customized memory controller that interfaces with the HD bitmap module and provides a way to allocate and access memory upon request. The HD bitmap module of a specific connection can request "set bit 5 of bitmap number 2", and the controller will know where in the memory bitmap number 2 of the given connection lies, it will access and change its 5th bit.

Each connection's HD bitmap comprises multiple equal-sized bitmap blocks, allocated as the need arises and exist at potentially different memory locations. Therefore, a connection must maintain a list of addresses, each pointing to the bitmap block. The main job of the memory controller is to efficiently store all these addresses for each connection.

A naive memory controller would maintain an array of addresses for each connection. The array indices refer to the subsequent bitmap block numbers while their values correspond to the physical memory addresses of those bitmap blocks. However, this approach requires memory-heavy state-keeping. For example, if we cap the total size of the HD bitmap to 256 bits and each bitmap block is 16 bits, then a connection can have a maximum of 16 bitmap blocks and would need to maintain the memory address for all of them. This is not desirable and leaves a huge memory footprint when considered on top of the states we need to maintain for the inner workings of the HD bitmap (mentioned in Table \ref{tab:states}). [for the rest of this section, we use this bitmap and block size for calculations related to HD bitmap].

A better approach would be to only store the absolute address of the first bitmap block and store the relative addresses of the subsequent blocks. Say the first block's address is $base\_bitmap\_address$, for all subsequent blocks allocated, the controller calculates the difference between the new block's address and $base\_bitmap\_address$, and stores that number in the array. To access any bitmap block $x>0$, the controller calculates its exact address via $base\_bitmap\_address$ + (value at index x). Now, for the same example above, rather than storing 16 addresses, the controller stores 1 address and 15 numbers which can be accommodated in fewer bytes than storing a memory address (2 bytes per number/address for our example).

\begin{figure}[t]
    \centering
     \includegraphics[width=\linewidth]{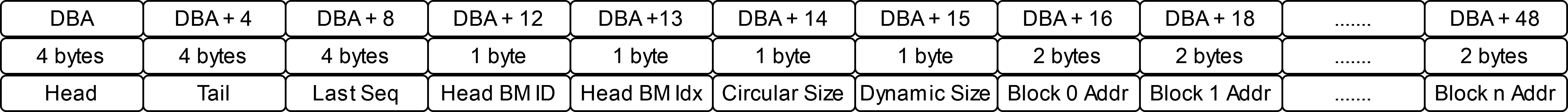}
    \caption{State-keeping by Memory Controller for one Connection}
    \label{fig:mem_controller_main}
    \vspace{-2mm}
\end{figure}

While the above approach halves the memory footprint of storing bitmap addresses, it is important to notice that we need to maintain this for each connection, on top of the states mentioned in Table \ref{tab:states}. In total for each connection we need to store 50 bytes worth of states, 16 bytes for variables, and 34 bytes for dynamic bitmap addresses of a 256-bit bitmap with a 16-bit block, even if the connection does not require an HD bitmap. To avoid this, we move these states inside the memory controller which only allocates memory to store these states for connections that face packet reordering (and require HD bitmap). Now each connection only maintains a Dynamic BM Address (DBA), which is initialized to NULL. If an out-of-order packet is received, the HD bitmap module requests the memory controller to initialize its states, which is done as shown in Fig \ref{fig:mem_controller_main}. Where the first 16 bytes are used for storing state variables from Table \ref{tab:states}, and following for HD bitmap addressing. Now the memory controller can help the module access all the necessary states and bitmaps since it knows the absolute address of states and the relative address of the bitmaps for each connection. This approach results in a lower amortized memory footprint (\S\ref{sec:implementation}).

\subsubsection{Trade offs: } There exists a trade-off between the memory needed to store the relative address of bitmap blocks vs the size of the one bitmap block. In our example above, for a maximum bitmap size of 256 bits, a 16-bit block size would result in 16 bitmap blocks needing 16*2 bytes for storing relative addresses. If we increase the block size to 32-bit or 64-bit, the relative addressing would require 8*2 or 4*2 bytes, respectively. However, a larger block size would also mean that even for connections facing little reordering a bigger space would be allocated to its bitmap. Although, if a network is bound to face a high degree of reordering then setting a larger block size would be more useful.

\begin{wraptable}{r}{0.6\textwidth}
\vspace{-4mm}
  \caption{Memory footprint vs OOO degree handled / connection}
  \vspace{-2mm}
  \label{tab:comparison}
  \small
  \resizebox{0.6\textwidth}{!}{
  \begin{tabular}{|c|c|c|c|}
    \hline
    \textbf{Name} & \textbf{Bitmap Memory} & \textbf{Other Data} & \textbf{Max OOO Degree Handled} \\
    \hline
    IRN & 125 bytes (100G links, 8us RTT) & 20 bytes & 1 BDP (100 packets) \\
    \hline
    MPRDMA & 17 bytes & 49 bytes & 64 packets \\
    \hline
    Eunomia & 4 - 82 bytes & 0 bytes & 0-256 packets \\
    \hline
  \end{tabular}
  }
  \vspace{-4mm}
\end{wraptable}

\subsubsection{Existing Schemes Comparison}
Compared to existing schemes IRN \cite{mittal2018revisiting} and MPRDMA \cite{lu2018multi}, Eunomia incurs a lower amortized memory footprint, and even when comparable to existing techniques, it handles a high degree of reordering for the same memory footprint, a brief comparison is provided in Table \ref{tab:comparison}.

\subsection{Receiver-Side Agent}
Apart from housing the HD bitmap and memory controller, the receiver-side is responsible of some other operations detailed below.

\subsubsection{Generating Acknowledgments}
After passing the packet through the HD bitmap, the receiver side generates either an ACK for in-order delivery, a SACK for OOO delivery where the packet is tracked in the bitmap, or NACK where a packet could not be tracked in the bitmap and consequently dropped. As mentioned earlier, these acknowledgments are not part of the congestion control running on the NICs and only add $\approx$5\% bandwidth overhead (for experiments in \S\ref{sec:eval_fine_grain_lb}).

\subsubsection{In-order delivery to Application}

Eunomia places the data directly in the application memory, regardless of the order it is received as long as it is tracked in the bitmap. It is important to ensure the application does not start reading data before it is fully received in order. For different RDMA verbs, this operation is performed differently:

\begin{compactitem}
    \item SEND/RECV: These RDMA verbs are two-way connections, communicating via pre-established send and receive queue pair (QP). Eunomia ensures in-order delivery to the application by holding off the completion notification (CN) until all the packets up to the \textit{Last Seq} have been received.
    \item WRITE: It is a one-sided verb and does not rely on CN by the NIC, rather the applications are responsible for synchronizing among each other. This is often done by placing a synchronization flag in the last packet, indicating the receiver application to start reading the data. Such a design makes it challenging to ensure the application does not read OOO data because if the last packet is received OOO, it can trigger the application to read wrong/incomplete data. To this end, Eunomia handles such a case by delaying the last packet (with synchronization) at the sender until it has received acknowledgment that all data till the last packet has been received in order. This approach however puts a latency overhead of 1 RTT, and we demonstrate this overhead in the appendix with potential improvement suggestions.
\end{compactitem}

\subsubsection{Garbage Collection}

The HD bitmaps are needed as long as the connection is active, once finished, the space should be freed up. Eunomia incorporates garbage collection to achieve this. Once everything up to the $Last Seq$ is received in order, the memory controller de-allocates and resets the HD bitmap and its states.

\section{Implementation}
\label{sec:implementation}

As Eunomia preserves the core structure of the original RoCEv2 protocol, a complete protocol rewrite is not necessary. Instead, we introduce a reordering layer on the receiver side to manage packet reordering effectively.

\begin{figure}[t]
    \begin{minipage}[b]{0.48\linewidth}
        \centering
        \begin{subfigure}[b]{0.49\linewidth}
            \centering
            \includegraphics[width=0.8\linewidth]{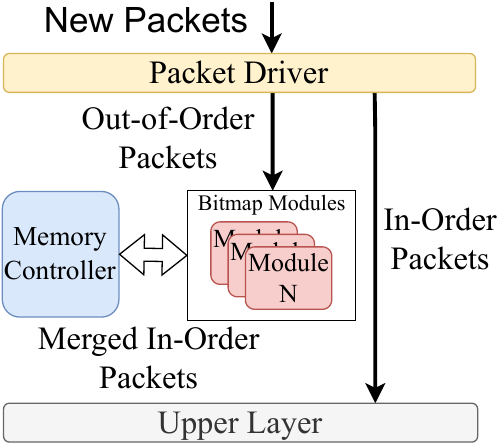}
            \caption{FPGA Block Design}
            \label{fig:fpga_block_design}
        \end{subfigure}
        \hfill
        \begin{subfigure}[b]{0.49\linewidth}
            \centering
            \includegraphics[width=0.7\linewidth]{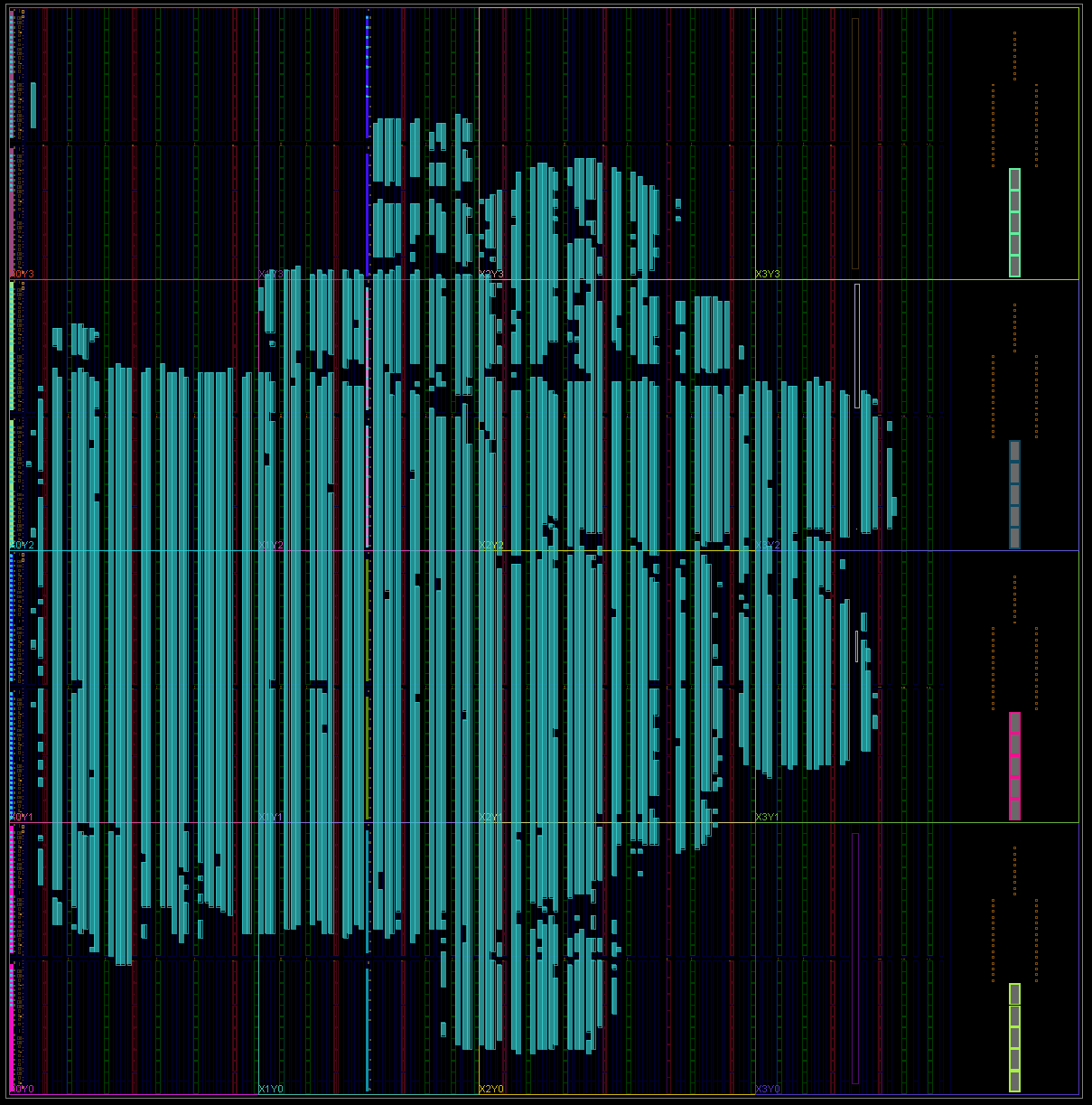}
            \caption{FPGA Chip Utilization}
            \label{fig:fpga_chip_util}
        \end{subfigure}
        \caption{FPGA Design and Utilization}
        \label{fig:fpga_design_utilization}
    \end{minipage}
    \hfill
    \begin{minipage}[b]{0.48\linewidth}
        \centering
        \begin{subfigure}[t]{\linewidth}
            \centering
            \includegraphics[width=0.6\linewidth]{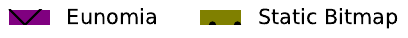}
        \end{subfigure}
        \begin{subfigure}[b]{0.49\linewidth}
            \centering
            \includegraphics[width=0.9\linewidth]{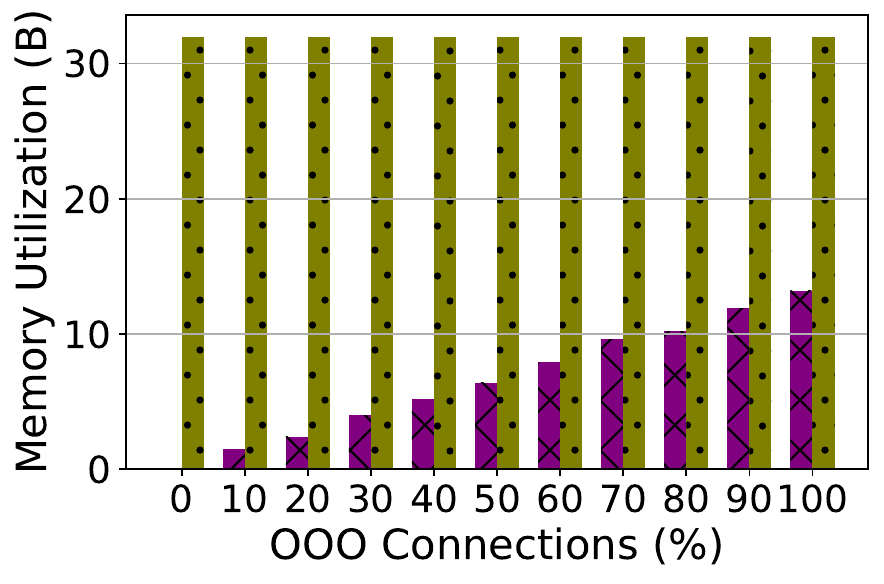}
            \caption{Bitmap Memory}
            \label{fig:mem_util_bitmap}
        \end{subfigure}
        \hfill
        \begin{subfigure}[b]{0.49\linewidth}
            \centering
            \includegraphics[width=0.9\linewidth]{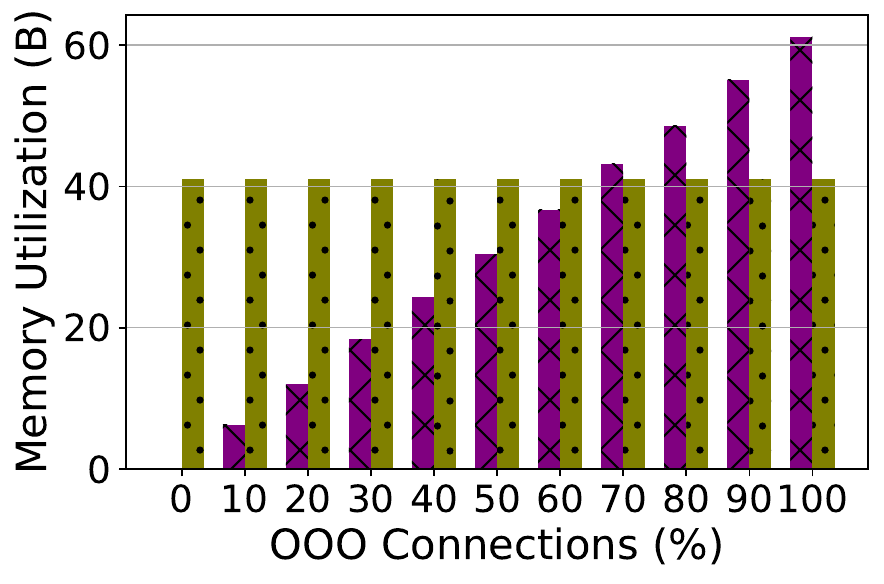}
            \caption{Total Memory}
            \label{fig:mem_util_total}
        \end{subfigure}
        \caption{Average Memory Utilization Per Connection}
        \label{fig:memory_utilization}
    \end{minipage}
    \vspace{-4mm}
\end{figure}

\paragraph{\textbf{Overview}}
Eunomia incorporates three core components: the HD bitmap module, the memory controller module, and the packet driver module. Together, these modules function as a reordering layer complementing the existing RoCEv2 protocol, as shown in Figure~\ref{fig:fpga_block_design}. Eunomia supports the deployment of multiple HD bitmap modules to accommodate multiple concurrent connections. In our current implementation, Eunomia has been tested with one packet driver, one memory controller, and a single HD bitmap module. Table~\ref{tab:fpga_utilization} presents the FPGA resource breakdown for this setup, while Figure ~\ref{fig:fpga_chip_util} illustrates the on-chip hardware consumption. 

\paragraph{\textbf{Test Platform}}
All modules are implemented in SystemVerilog, and we simulate our design using an AMD Kintex UltraScale+ FPGA KCU116 base model, specifically the Kintex UltraScale+ XCKU5P-2FFVB676E FPGA, with Vivado 2022.2. However, our design is not tied to a specific FPGA board and is adaptable to any FPGA capable of supporting RoCEv2 operations.

\paragraph{\textbf{Packet Driver}}
The packet driver is the entry point of any incoming packets. It determines whether a packet is out-of-order and invokes the HD bitmap module when needed. It has two critical components 
\begin{compactitem}
    \item \textbf{conn\_to\_module\_map:} This array is responsible for mapping an active connection to a bitmap module. A record will only be entered once an out-of-order packet arrives for the first time.
    \item \textbf{conn\_module\_valid:} This array indicates whether there is a current active bitmap module for a connection.
\end{compactitem}

\begin{wraptable}{r}{0.6\textwidth}
\vspace{-4mm}
  \caption{FPGA Utilization Breakdown}
  \vspace{-2mm}
  \label{tab:fpga_utilization}
  \small
  \resizebox{0.6\textwidth}{!}{
  \begin{tabular}{|c|c|c|c|c|c|c|}
    \hline
    \textbf{Name} & \textbf{LUT} & \textbf{CARRY8} & \textbf{Registers} & \textbf{F7 Muxes} & \textbf{F8 Muxes}  \\
    \hline
    Eunomia Total        & 82201 & 792 & 3434 & 3800 & 1004\\
    \hline
    Memory Controller    & 81029 & 521 & 2351 & 3429 & 1001 \\
    \hline
    HD Bitmap      & 2180  & 253 & 613  & 34   & 0  \\
    \hline
  \end{tabular}
  }
  \vspace{-4mm}
\end{wraptable}

\paragraph{\textbf{HD Bitmap}}
The HD bitmap module contains the logic of tracking out-of-order packets in the bitmap and is only activated for connections facing reordering. The packet driver assigns a connection ID (of the current connection) to the bitmap module each time the connection state is updated (e.g. when a new packet arrives or a timeout occurs). The HD bitmap module uses this connection ID to retrieve and update relevant bitmap data from the memory controller. To enable module sharing and memory preservation, the HD bitmap module itself does not store any connection-specific data, instead, all storage, including metadata such as head, tail, and last sequence, is delegated to the memory controller.

\paragraph{\textbf{Memory Controller}}
The memory controller is responsible for storing all connection-related data and contains three primary storage arrays:
\begin{compactitem}
    \item \textbf{master\_array:} This main block is responsible for storing all metadata and bitmap information for each connection. It occupies the available memory, aside from the portions allocated to the other two arrays. The master array is partitioned into 2-byte blocks, with a current configurable capacity of 1024 blocks.
    \item \textbf{block\_alloc\_bitmap:} This allocation bitmap tracks the allocation status of each block within the master array, which marks blocks that are currently in use. It is a one-bit array with a length equal to the number of blocks in the master array (1024 in the current design).
    \item \textbf{metadata\_start\_index:} This fixed-length array stores the starting index of the metadata for each connection. Currently, each entry occupies 3 bytes: 1 byte for the connection ID and 2 bytes for the starting index.
\end{compactitem}

\textbf{Finding New Space: }
When a connection requires a new bitmap block, the memory controller searches for the first available block, beginning from the head of the master array, as indicated by the block allocation bitmap. When a new connection is instantiated, the memory controller searches the master array backward, from the tail to the head for a sequence of consecutive available blocks that can accommodate all metadata for the connection (24 blocks in our current design). In the future, we may introduce an alternative structure, such as a linked list, to enable storage of metadata in non-consecutive blocks.

\textbf{Reset: }
When a connection is completed or terminated, the memory controller de-allocates all associated blocks by updating the block\_alloc\_bitmap. The actual data within the master array is not erased but will be overwritten when a new connection is initialized in that space.

\paragraph{\textbf{Memory Utilization}}
To test Eunomia, we conducted experiments designed to reflect varying conditions of packet reordering and compared Eunomia's memory utilization against a naive static bitmap implementation. In Eunomia, the maximum bitmap size is set to 256 bits with a 16-bit block size. For a fair comparison, the size of the static bitmap is set to 256 bits (32 bytes) so it can handle the same amount of reordering as Eunomia. The simple static bitmap would require an additional 9 bytes to function, 4 bytes each for the head and last packet and 1 byte for the head index, totaling 41 bytes per connection. Each experiment comprises 20 concurrent connections, with 0\% to 100\% of the connections facing reordering, with the degree of reordering varying between 10\% and 100\%.

The results of these experiments, presented in Fig ~\ref{fig:memory_utilization}, Fig ~\ref{fig:mem_util_bitmap} reveals that Eunomia consistently achieves lower memory utilization for bitmaps alone compared to the static bitmap approach. Eunomia’s dynamic allocation of bitmap blocks effectively minimizes unused space, which contrasts sharply with the fixed-size design of the static bitmap. Memory is consumed by Eunomia’s bitmap module only when OOO packets occur, demonstrating significant memory efficiency, especially in scenarios with low to moderate OOO traffic.

It is important to note that Eunomia’s overall memory utilization includes an additional 48 bytes per connection for metadata. So, as the OOO percentages exceed 60\%, Eunomia’s average memory utilization surpasses that of the static bitmap (Fig \ref{fig:mem_util_total}). This increase is attributed to the relatively higher proportion of metadata in Eunomia’s memory profile under the current setup. However, with a larger block size the impact of metadata on total memory usage becomes proportionally smaller. Furthermore, it is important to contextualize these results within real-world applications. Systems employing static bitmap approaches, such as MPRDMA \cite{lu2018multi} and IRN \cite{mittal2018revisiting}, typically require more than 41 bytes per connection and handle lower degree of reordering compared to Eunomia (\S\ref{sec:memory_controller}). Thus, Eunomia presents a compelling advantage in scenarios where robust OOO handling is critical. It offers efficient memory utilization under most practical conditions and significantly reduces memory overhead under a high degree of packet reordering.

\section{Evaluation}
\label{sec:eval}
In this section, we compare Eunomia with existing schemes, using different performance metrics, such as flow completion time, throughput, and PFC Pause duration. We also run experiments for micro-benchmarking of Eunomia in terms of memory footprint. We run our experiments in ns-3 simulations \cite{ns3_simul}.

\textbf{Common Setup: } Unless specified otherwise, in all our experiments, the topology used is an 8x8 Clos with 128 hosts (1:2 over-subscription), and the links are 100G with 1us propagation delay. The switches use a shared buffer architecture, with a size set to 9MB and PFC enabled. All hosts use DCQCN as congestion control, the parameters are set based on \cite{song2023network}.

\textbf{Network Traffic}: We generate traffic using MetaHadoop distribution \cite{roy2015inside}, which provides heavy-tailed traffic, and Ali Storage distribution \cite{li2019hpcc}, which provides higher variability in flow sizes. We use the Poisson distribution to calculate the inter-arrival times between flows and change the inter-arrival times to vary the network load.

\textbf{Baselines: } To test Eunomia we pair it with different existing techniques, including packet spray (PS) LB, DRILL LB, Power of 2 (PO2) LB, shortest-remaining-processing-time (SRPT) scheduling, and simple deflection (DIBS). We compare them against ECMP and Conweave (parameters set based on \cite{song2023network}) with PFC or IRN where applicable. We omit MP-RDMA from our evaluation comparison because of its sensitivity to parameter tuning and we observed that it performs worse than Conweave in Fig \ref{fig:motivation_clos_avg_fct_80p}, so we only compare our schemes against Conweave from existing work.

The key findings of our experiments are:
\begin{compactitem}
    \item Eunomia enables RDMA networks to leverage a variety of existing performance-enhancing techniques such as fine-grained LBs, irregular topologies, sophisticated flow scheduling, failure management, and incast management.
    \item With Eunomia enabled RDMA is able to significantly reduce flow completion times by up to $91\%$ and $55\%$, and PFC triggers by by $\approx99\%$ compared to existing state-of-the-art schemes.
\end{compactitem}

\subsection{Eunomia enables fine-grained load balancers in RDMA Datacenters}
\label{sec:eval_fine_grain_lb}

Enabling ordering support using Eunomia on the NIC allows fine-grained LBs to significantly reduce flow-completion times compared to existing techniques such as ECMP, Conweave, and IRN.

\begin{figure}[t]
  \centering
  \vspace*{-\baselineskip}
  \begin{subfigure}{\linewidth}
    \centering
    \includegraphics[width=0.7\linewidth]{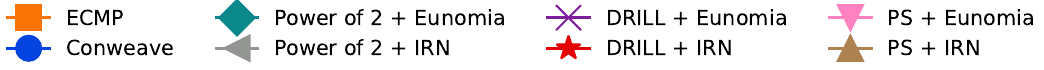}
  \end{subfigure}

  \begin{subfigure}{0.24\linewidth}
    \centering
    \includegraphics[width=\linewidth]{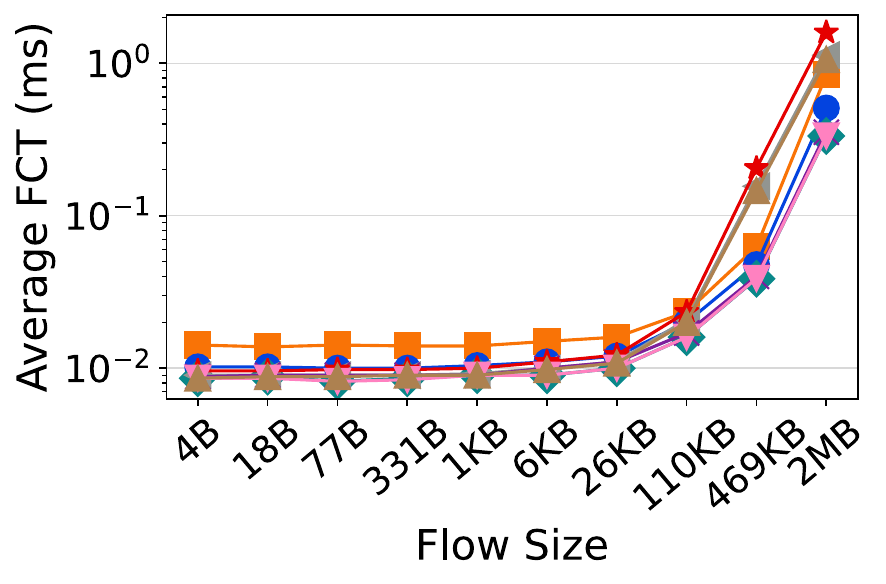}
    \caption{Avg FCT at 50\% Load}
    \label{fig:50_avg_fct_slowdown_ali}
  \end{subfigure}
  \hfill
  \begin{subfigure}{0.24\linewidth}
    \centering
    \includegraphics[width=\linewidth]{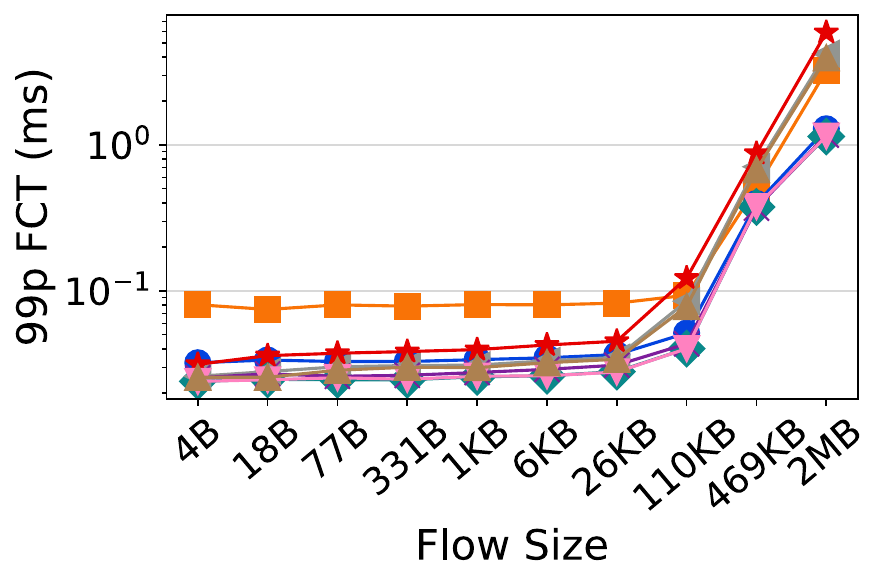}
    \caption{Tail FCT at 50\% Load}
    \label{fig:50_tail_fct_slowdown_ali}
  \end{subfigure}
  \hfill
  \begin{subfigure}{0.24\linewidth}
    \centering
    \includegraphics[width=\linewidth]{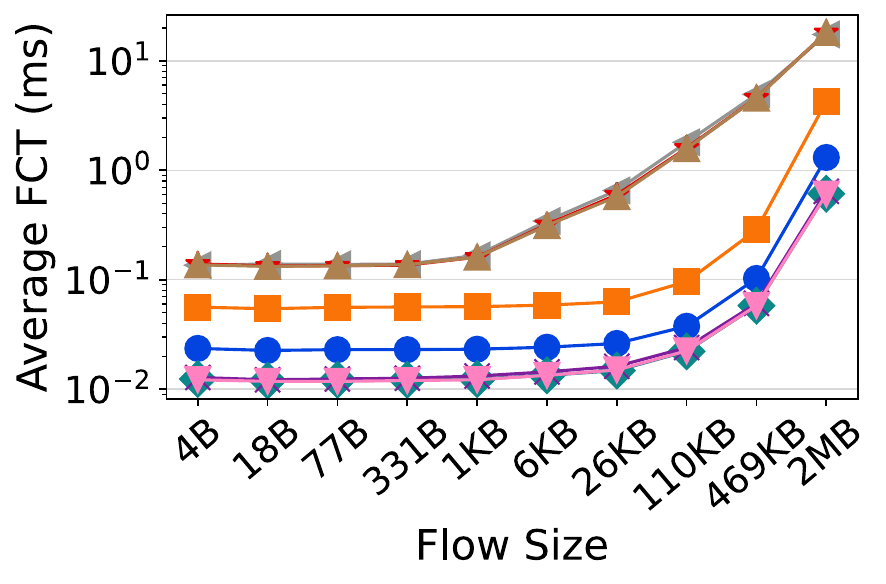}
    \caption{Avg FCT at 80\% Load}
    \label{fig:80_avg_fct_slowdown_ali}
  \end{subfigure}
  \hfill
  \begin{subfigure}{0.24\linewidth}
    \centering
    \includegraphics[width=\linewidth]{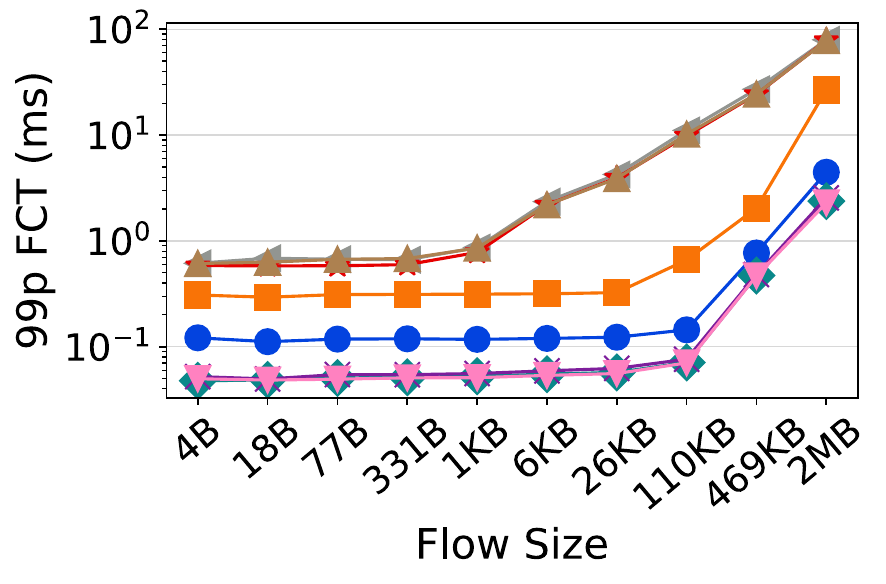}
    \caption{Tail FCT at 80\% Load}
    \label{fig:80_tail_fct_slowdown_ali}
  \end{subfigure}
  \vspace{-2mm}
  \caption{Eunomia achieves significant improvement in (a) (c) average and (b) (d) tail flow completion times in RDMA networks compared to state-of-the-art techniques for AliStorage workload.}
  \label{fig:clos_fct_slowdown}
  \vspace{-2mm}
\end{figure}

Eunomia successfully handles the reordering induced by fine-grained LBs. Fig \ref{fig:clos_fct_slowdown} shows the mean and tail FCTs of different flow sizes under moderate and high network load. Eunomia aids these LBs in reducing FCT in comparison with existing techniques such as ECMP and Conweave. In Fig \ref{fig:50_tail_fct_slowdown_ali} we observe that under moderate network load Eunomia reduces the tail FCT by up to $\approx70\%$ and $\approx26\%$ for short flows and up to $\approx65\%$ and $\approx21\%$ for long flows, compared to ECMP and Conweave respectively. At a high network load in Fig \ref{fig:80_avg_fct_slowdown_ali} and \ref{fig:80_tail_fct_slowdown_ali}, the performance gains are significantly larger. Compared to ECMP and Conweave, finer-grained LBs with Eunomia improve mean FCT by up to $\approx85\%$ and $\approx53\%$ and tail FCT by up to $\approx91\%$ and $\approx55\%$, respectively.

\begin{figure}[t]
  \centering
  \begin{subfigure}{\linewidth}
    \centering
    \includegraphics[width=\linewidth]{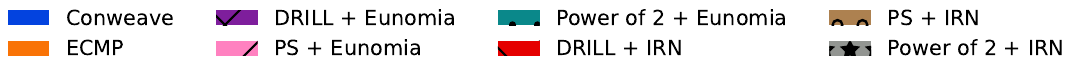}
  \end{subfigure}

  \begin{subfigure}{0.24\linewidth}
    \centering
    \includegraphics[width=\linewidth]{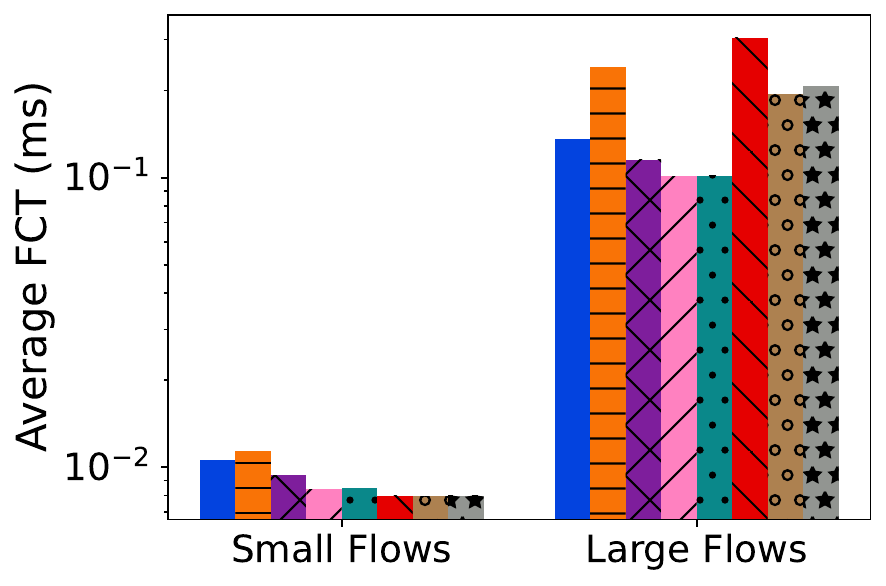}
    \caption{Avg FCT at 50\% Load}
    \label{fig:50_avg_fct_slowdown_hadoop}
  \end{subfigure}
  \hfill
  \begin{subfigure}{0.24\linewidth}
    \centering
    \includegraphics[width=\linewidth]{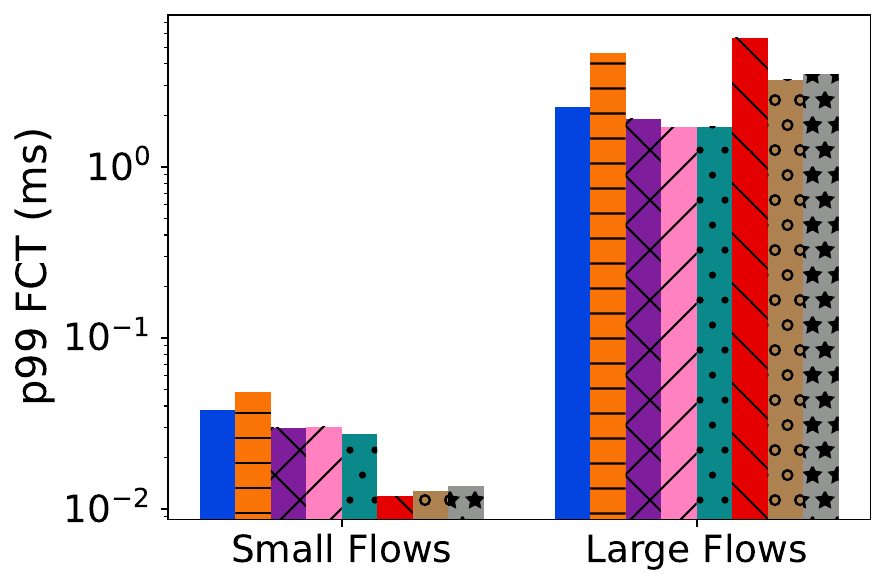}
    \caption{Tail FCT at 50\% Load}
    \label{fig:50_tail_fct_slowdown_hadoop}
  \end{subfigure}
  \hfill
  \begin{subfigure}{0.24\linewidth}
    \centering
    \includegraphics[width=\linewidth]{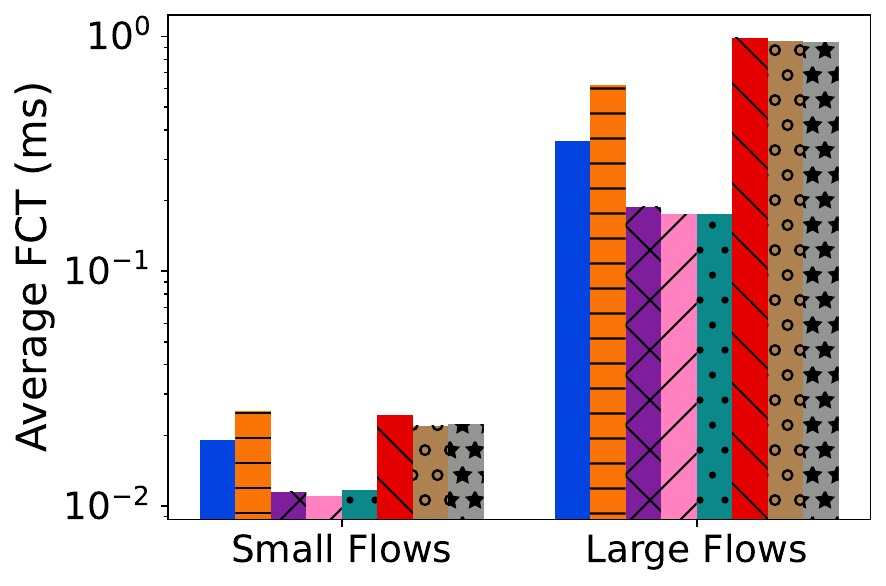}
    \caption{Avg FCT at 80\% Load}
    \label{fig:80_avg_fct_slowdown_hadoop}
  \end{subfigure}
  \hfill
  \begin{subfigure}{0.24\linewidth}
    \centering
    \includegraphics[width=\linewidth]{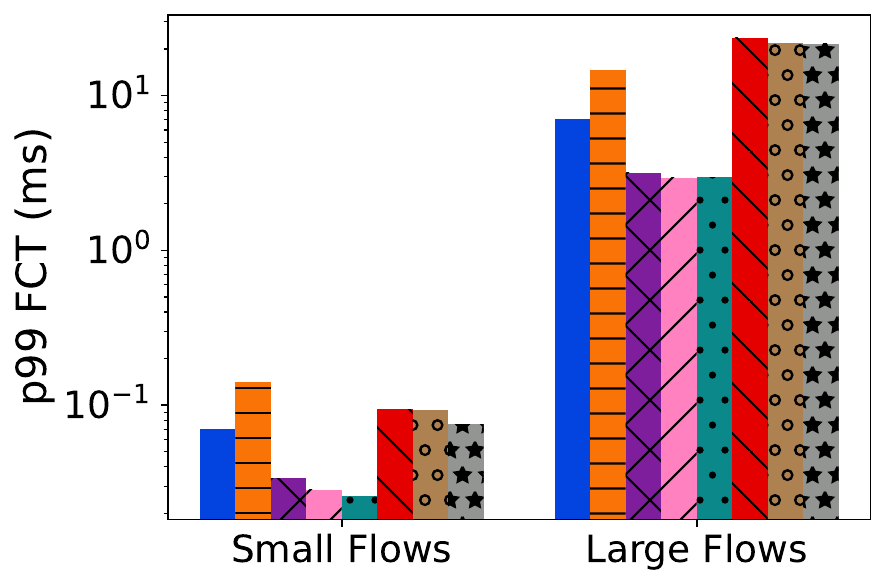}
    \caption{Tail FCT at 80\% Load}
    \label{fig:80_tail_fct_slowdown_hadoop}
  \end{subfigure}
  \caption{Eunomia improves (a) (c) average and (b) (d) tail flow completion times in RDMA networks compared to state-of-the-art techniques for FBHadoop workloads.}
  \label{fig:clos_fct_slowdown_hadoop}
  \vspace{-2mm}
\end{figure}

Fig \ref{fig:50_avg_fct_slowdown_hadoop} shows the results of the same setup with FBHadoop workload. We observe performance gains allowed by Eunomia for small flows ($\leq10KB$) and for large flows ($\geq 5000KB$). At moderate load, the tail FCT of small flows is improved by $\approx44\%$ and $\approx28\%$, while that of large flows is improved by $\approx24\%$ and $\approx63\%$ in fine-grained LB (with Eunomia) against ECMP and Conweave (Fig \ref{fig:50_tail_fct_slowdown_hadoop}). At high network load, the performance gains of fine-grained LBs over ECMP and Conweave are more pronounced (Fig \ref{fig:80_avg_fct_slowdown_hadoop}), improving FCT by upto $\approx63\%$ and $\approx82\%$, respectively.

\subsection{Eunomia helps reduce the PFC triggers}
\label{sec:eval_pfc_triggers}
As a bi-product of leveraging multiple paths in the network and efficient handling packet reordering, Eunomia significantly lowers PFC triggers and their duration compared to existing techniques.

\begin{figure}[t]
  \centering
  \begin{subfigure}{\linewidth}
    \centering
    \includegraphics[width=\linewidth]{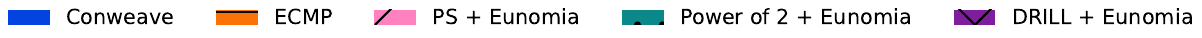}
  \end{subfigure}

  \begin{subfigure}{0.24\linewidth}
    \centering
    \includegraphics[width=\linewidth]{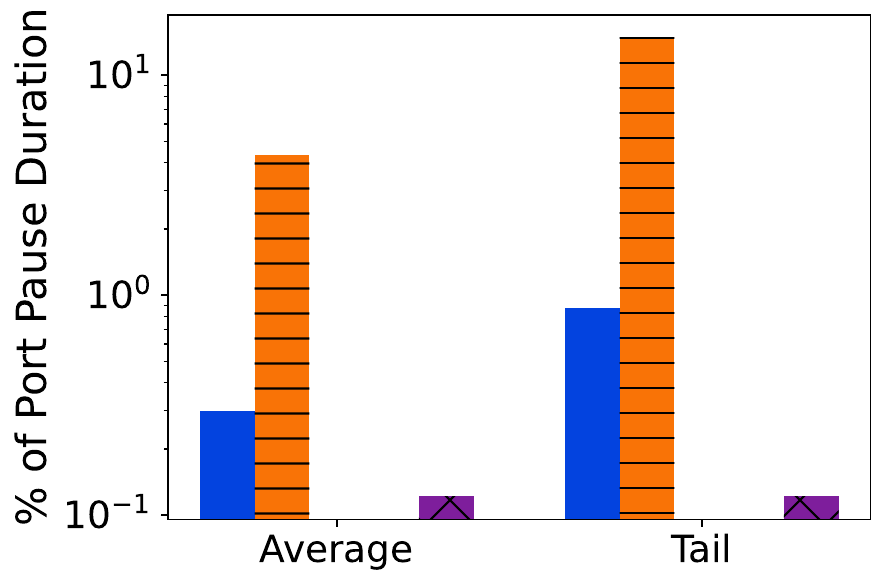}
    \caption{Switch at 50\% Load}
    \label{fig:50_port_pause_switch}
  \end{subfigure}
  \hfill
  \begin{subfigure}{0.24\linewidth}
    \centering
    \includegraphics[width=\linewidth]{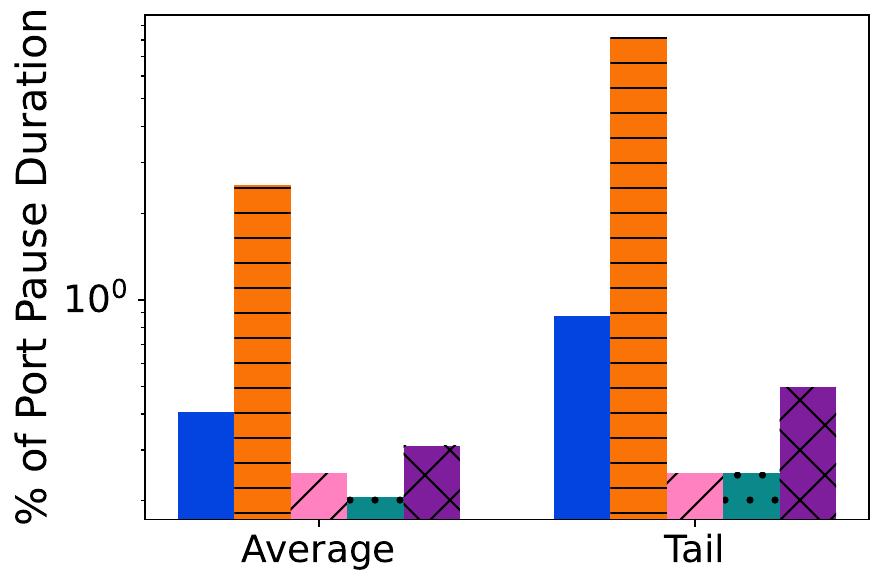}
    \caption{Server at 50\% Load}
    \label{fig:50_port_pause_server}
  \end{subfigure}
  \hfill
  \begin{subfigure}{0.24\linewidth}
    \centering
    \includegraphics[width=\linewidth]{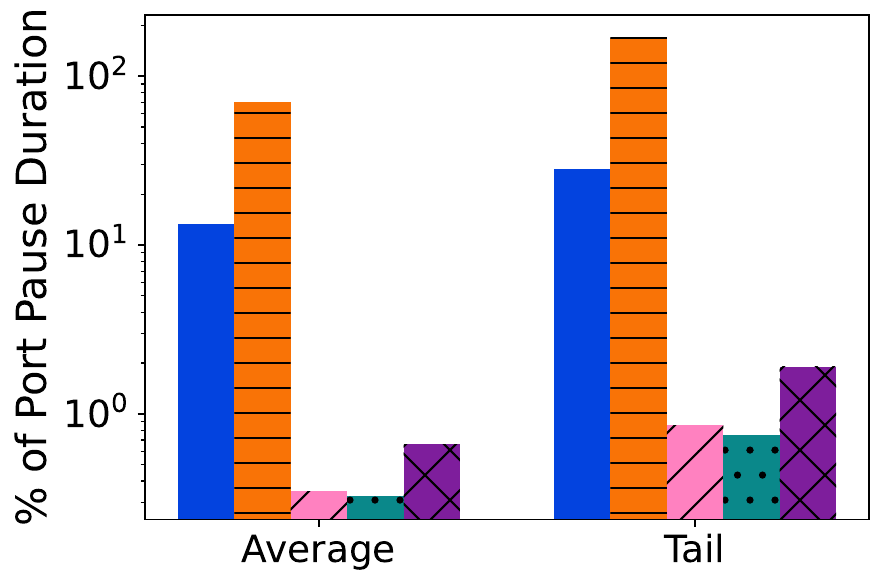}
    \caption{Switch at 80\% Load}
    \label{fig:80_port_pause_switch}
  \end{subfigure}
  \hfill
  \begin{subfigure}{0.24\linewidth}
    \centering
    \includegraphics[width=\linewidth]{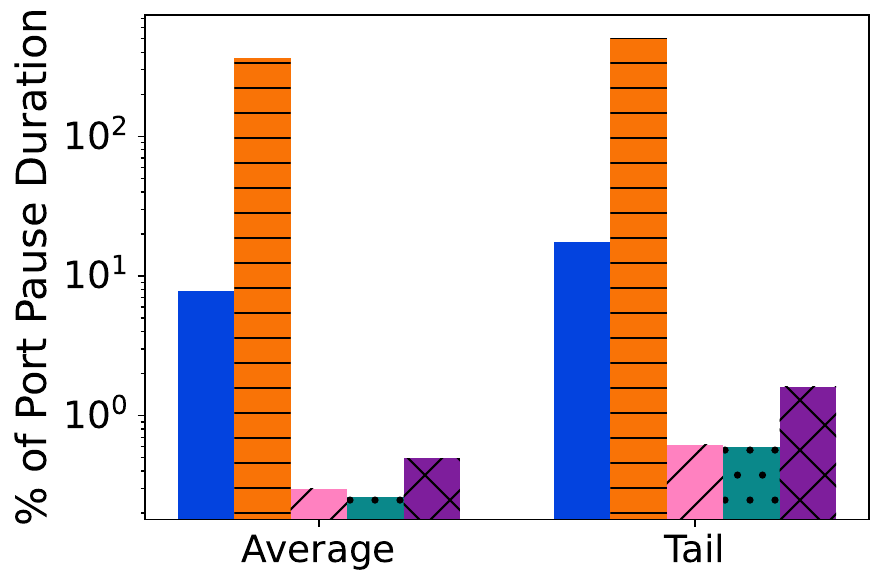}
    \caption{Server at 80\% Load}
    \label{fig:80_port_pause_server}
  \end{subfigure}

  \caption{Eunomia reduces the PFC pause duration on (a) (c) switch and (b) (d) server ports}
  \label{fig:pfc_clos_ali}
  \vspace{-2mm}
\end{figure}

Priority Flow Control (PFC) can negatively impact performance if triggered frequently \cite{guo2016rdma}. So, we evaluate the impact of Eunomia on PFC, Fig \ref{fig:pfc_clos_ali} shows the average fraction of the total time a port was paused in the network under moderate and high load. We observe that pause duration is almost non-existent in the case of finer-grained LBs paired with Eunomia. The presence of ordering support in RDMA allows efficient load distribution on multiple paths, which reduces hot spots in the network, and consequently reduces the chance of PFC getting triggered on those spots, ECMP and Conweave on the other hand face PFC pause for a longer duration consequently degrading in performance.

It is also worth noticing in Fig \ref{fig:50_port_pause_server} and Fig \ref{fig:80_port_pause_server} that servers, as opposed to switches, face higher pause duration in ECMP and Conweave. This happens because, in coarse-grained LBs, the chances of PFC spreading to uplink ports are higher since all or most packets are on the same (potentially congested) path. PFC can spread up to the servers and head-of-line block other flows that are not part of the congestion, resulting in degraded performance.

\subsection{Eunomia allows RDMA to be employed in irregular topolgies}
\label{sec:eval_irregular_topo}

RDMA can reap the benefits of irregular topologies like Jellyfish by incorporating Eunomia on the NICs, which enables multi-path routing, an important requirement of irregular topologies, to reduce FCT and increase throughput.

\textbf{Setup: } We use an 8-ary Fat tree topology and a corresponding jellyfish topology with the same number of servers and switches as the 8-ary Fat tree. This allows a fair comparison between both topologies. We use ECMP in Fat tree while in Jellyfish we use FKSP (to show a direct deployment of RDMA in Jellyfish that does not need ordering support) and Power of 2 along with Eunomia because fine-grained routing enables Jellyfish to deliver desired performance gains.

\begin{figure}[t]
  \centering
  \begin{subfigure}{\linewidth}
    \centering
    \includegraphics[width=0.8\linewidth]{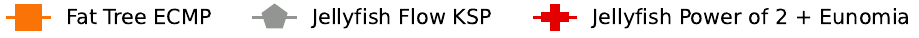}
  \end{subfigure}

  \begin{subfigure}{0.28\linewidth}
    \centering
    \includegraphics[width=\linewidth]{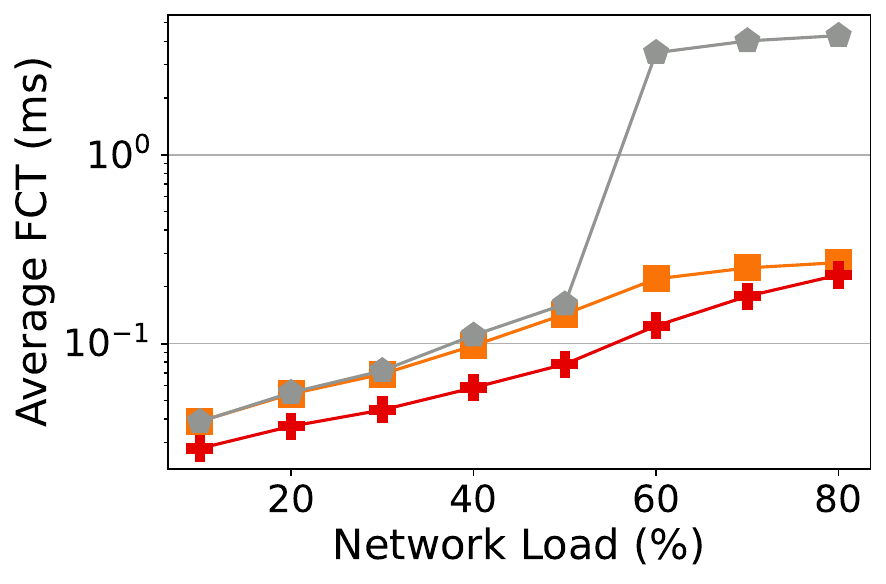}
    \caption{Avg FCT}
    \label{fig:avg_fct_jellyfish_all_loads}
  \end{subfigure}
  \hfill
  \begin{subfigure}{0.28\linewidth}
    \centering
    \includegraphics[width=\linewidth]{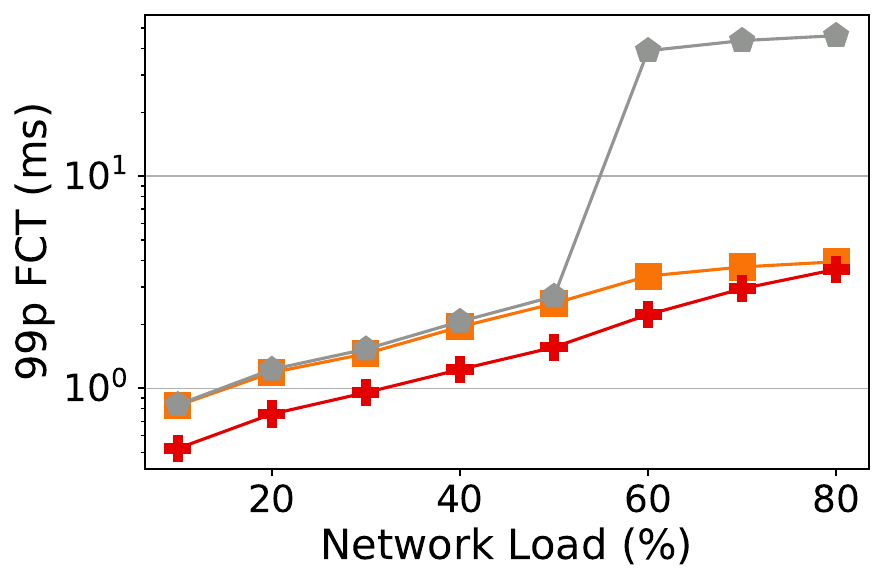}
    \caption{Tail FCT}
    \label{fig:tail_fct_jellyfish_all_loads}
  \end{subfigure}
  \hfill
  \begin{subfigure}{0.28\linewidth}
    \centering
    \includegraphics[width=\linewidth]{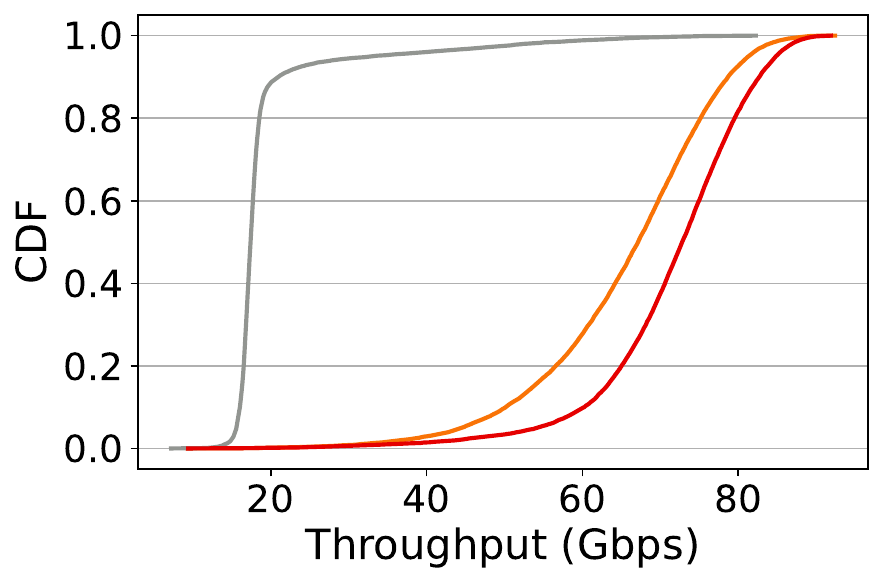}
    \caption{Throughput}
    \label{fig:thrpt_80p_jellyfish_eval}
  \end{subfigure}
  \vspace{-2mm}
  \caption{Eunomia enables RDMA to reap the benefits of Jellyfish topology, with improved flow completion times and throughput than Fat tree. 
  }
  \label{fig:jellyfish_fct_s_vs_l_eval}
\vspace{-2mm}
\end{figure}

\noindent\textbf{Result: } Fig \ref{fig:avg_fct_jellyfish_all_loads} and Fig \ref{fig:tail_fct_jellyfish_all_loads} show mean and tail FCT at varying network loads, respectively. 
FKSP in Jellyfish delivers similar performance as Fat-tree (+ ECMP) at lower loads but starts to degrade at higher loads, because as the network load increases so do the chances of collisions. If many flows collide on longer paths, they face performance loss. These experiments show that RDMA can not be directly employed in expander graph topologies, and requires ordering support. RDMA with Power of 2 + Eunomia performs significantly better, achieving up to $94\%$ lower mean FCT than FKSP. This is because Eunomia efficiently handles the packet reordering induced by the topology and the LB. Jellyfish with Eunomia also improves performance compared to Fat-tree topology as seen in Fig \ref{fig:avg_fct_jellyfish_all_loads} and \ref{fig:tail_fct_jellyfish_all_loads}, giving $30\%$ and $\approx21\%$ lower mean and tail FCT at 70\% Load.

\begin{figure}[t]
    
    \centering
  \begin{subfigure}{\linewidth}
    \centering
    \includegraphics[width=0.9\linewidth]{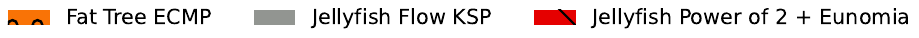}
  \end{subfigure}

  \begin{subfigure}{0.24\linewidth}
    \centering
    \includegraphics[width=\linewidth]{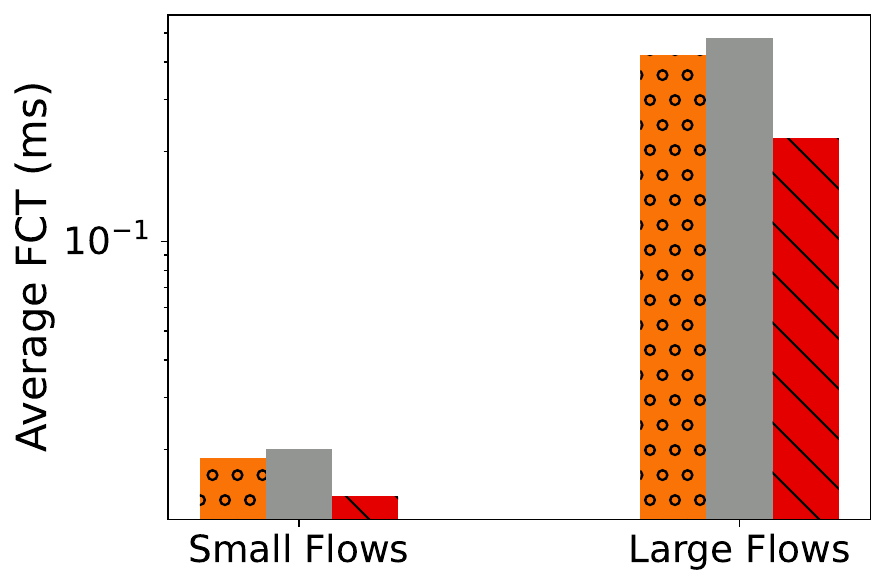}
    \caption{Avg FCT at 50\% Load}
    \label{fig:avg_fct_50p_jellyfish_eval}
  \end{subfigure}
  \hfill
  \begin{subfigure}{0.24\linewidth}
    \centering
    \includegraphics[width=\linewidth]{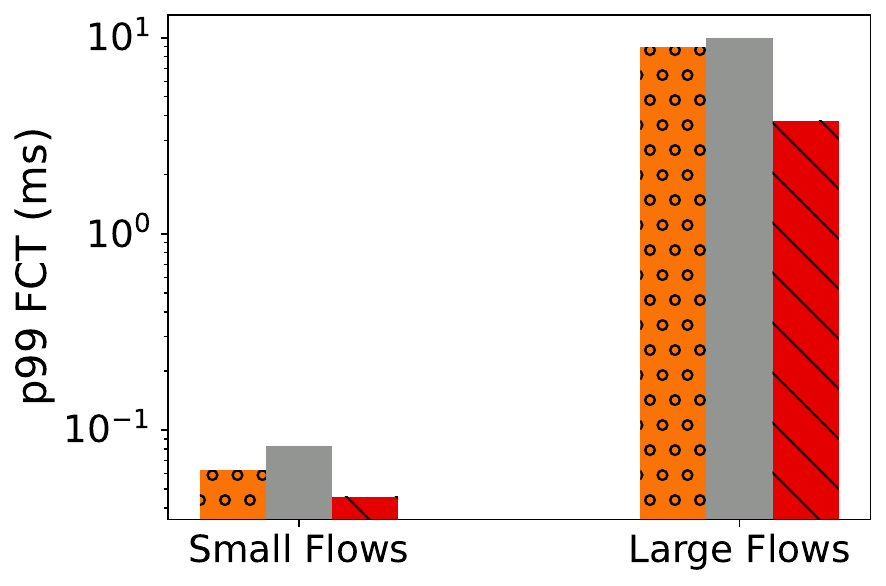}
    \caption{Tail FCT at 50\% Load}
    \label{fig:tail_fct_50p_jellyfish_eval}
  \end{subfigure}
  \begin{subfigure}{0.24\linewidth}
    \centering
    \includegraphics[width=\linewidth]{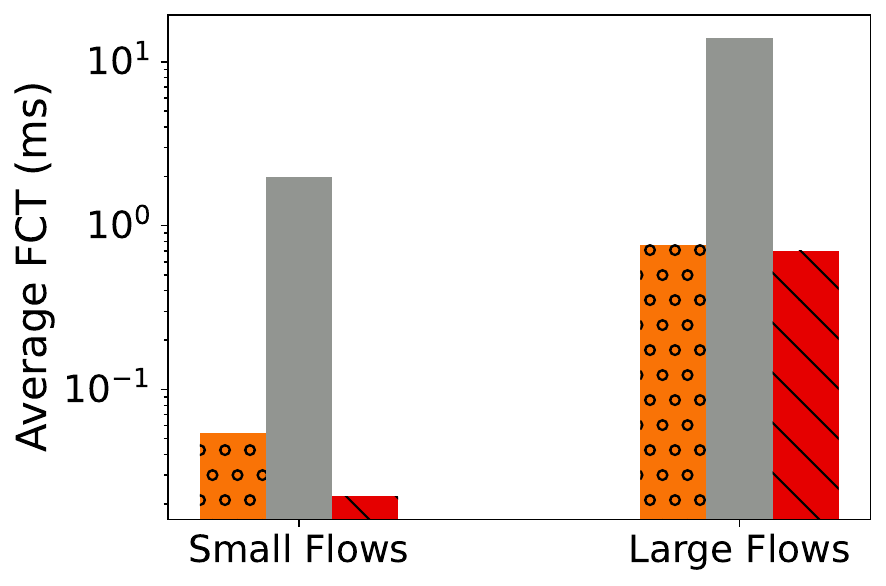}
    \caption{Avg FCT at 80\% Load}
    \label{fig:avg_fct_80p_jellyfish_eval}
  \end{subfigure}
  \hfill
  \begin{subfigure}{0.24\linewidth}
    \centering
    \includegraphics[width=\linewidth]{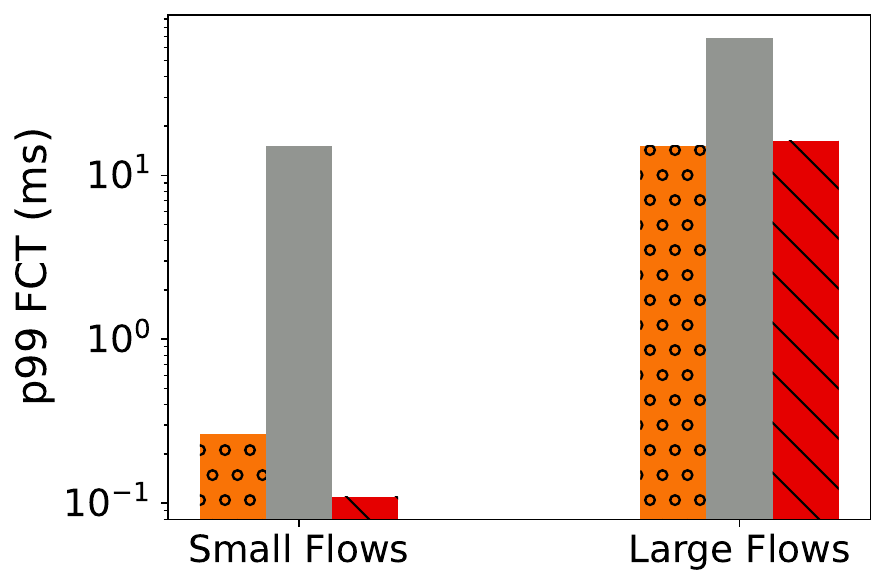}
    \caption{Tail FCT at 80\% Load}
    \label{fig:tail_fct_80p_jellyfish_eval}
  \end{subfigure}

  \caption{RDMA equipped with Eunomia in Jellyfish topologies improves the flow completion times of short flows at moderate and high load. 
  }
  \label{fig:jellyfish_fct_small_vs_large}    
\end{figure}

We zoom into the FCT at 50\% and 80\% load for different flow sizes in Fig \ref{fig:jellyfish_fct_small_vs_large}. We observe that at high network load (80\%), short flows (up to 16KB in size) perform $\approx60\%$ better in Jellyfish than in Fat-tree, moreover, the mean and tail FCT of large flows in Jellyfish are penalized because more flows reach completion in Jellyfish than in Fat-tree. This indicates that Eunomia allows Jellyfish topology to absorb more load.

In Fig \ref{fig:thrpt_80p_jellyfish_eval}, the aggregate throughput in Jellyfish with FKSP is quite low. But Eunomia enables Power of 2 to significantly improve throughput in Jellyfish, resulting in $2.7\times$ higher throughput than FKSP. It also allows RDMA to achieve higher throughput in Jellyfish than in Fat-tree.

\subsection{Eunomia enables flow scheduling in RDMA}
\label{sec:eval_flow_scheduling}

Eunomia enables RDMA to leverage sophisticated flow scheduling techniques such as shortest-remaining processing time (SRPT) \cite{alizadeh2013pfabric}, allowing significant performance gains for latency-sensitive, short flows.

\textbf{Setup: } For this experiment, we did not cap the HD bitmap in Eunomia. This allows Eunomia to handle reordering induced by SRPT scheduling for long flows, despite arriving in order at the TOR switch. We use two types of LBs here, ECMP since it is the state-of-art in RDMA, and PO2 because in our prior experiments, this has been the "best case" LB for this setup. We compare them both with first-in-first-out (FIFO) scheduling vs SRPT scheduling.

\textbf{Result: }Fig \ref{fig:srpt_vs_fifo} shows the performance of different flow sizes at 80\% load. The performance of short flows improves for both LBs using SRPT, up to 82\% (ECMP) and 86\% (PO2) lower average FCT compared to the default ECMP with FIFO in RDMA. Large flows, however, either do not improve (PO2) or worsen (ECMP) in SRPT than FIFO due to the inherent nature of SRPT to favor short flows. The degradation in ECMP is due to hash collisions, if many short and large flows collide on a path, SRPT will always prioritize short flows. It is important to notice that without capping the HD bitmap size in Eunomia, it grows significantly more than the FIFO case of the same LB, for example on average taking 200 bytes as opposed to 26 bytes. However, this increase is due to large flows arriving in reverse order due to SRPT, these flows can be opted out of this scheduling since they do not benefit from it.

\subsection{Eunomia enables Incast management via deflection}
\label{sec:eval_incast}

Eunomia enables deflection techniques, such as DIBS \cite{dibs} that allow efficient burst-absorption and reduce packet drops in the events of incast.

\textbf{Setup: } To handle incast we use simple deflection like DIBS \cite{dibs} paired with Eunomia and with IRN, we also use IRN alone as a baseline because incast is known to cause packet drops and IRN is made with a lossy network in mind. We experiment with varying degrees of incast traffic load, where the incast scale is 50 to 1 and each flow is 40KB.

\begin{figure}[t]
    \begin{minipage}[b]{0.48\linewidth}
        \centering
        \begin{subfigure}[b]{1.0\linewidth}
            \centering
            \includegraphics[width=1.0\linewidth]{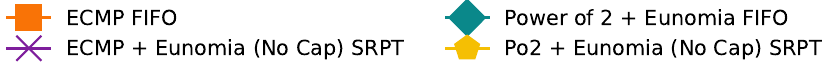}
            \label{fig:srpt_legend}
        \end{subfigure}
        \begin{subfigure}[b]{0.49\linewidth}
            \centering
            \includegraphics[width=\linewidth]{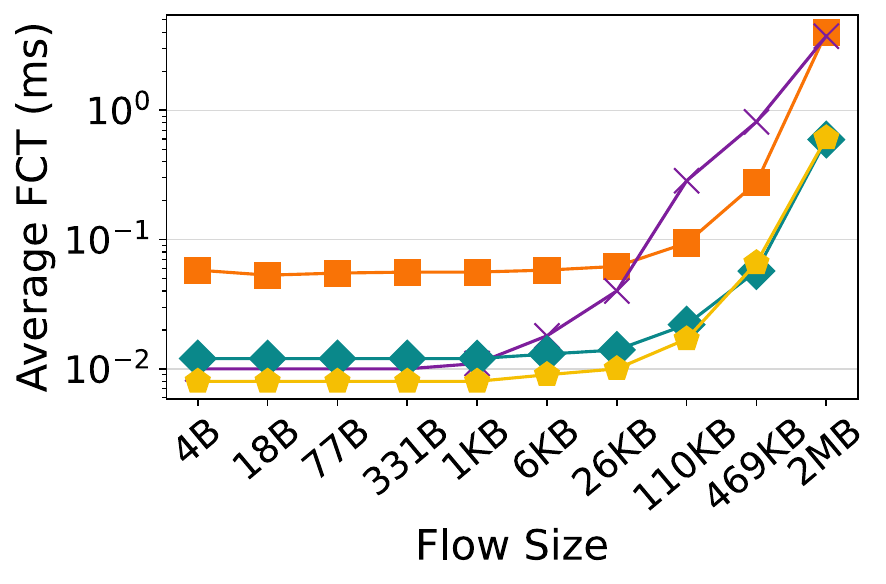}
            \caption{Average FCT}
            \label{fig:avg_fct_srpt_vs_fifo}
        \end{subfigure}
        \hfill
        \begin{subfigure}[b]{0.49\linewidth}
            \centering
            \includegraphics[width=\linewidth]{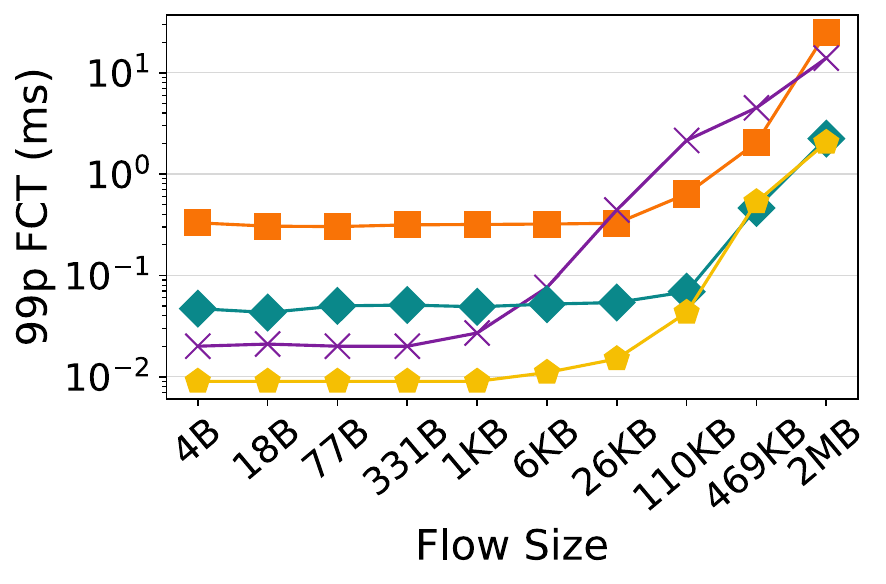}
            \caption{Tail FCT}
            \label{fig:tail_fct_srpt_vs_fifo}
        \end{subfigure}
        \vspace{-4mm}
        \caption{Eunomia enables SRPT scheduling in RDMA}
        \label{fig:srpt_vs_fifo}
    \end{minipage}
    \hfill
    \begin{minipage}[b]{0.48\linewidth}
        \begin{subfigure}[b]{0.6\linewidth}
            \centering
            \includegraphics[width=1.0\linewidth]{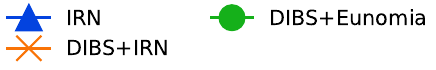}
        \end{subfigure}
        \begin{subfigure}{0.49\linewidth}
            \centering
            \includegraphics[width=\linewidth]{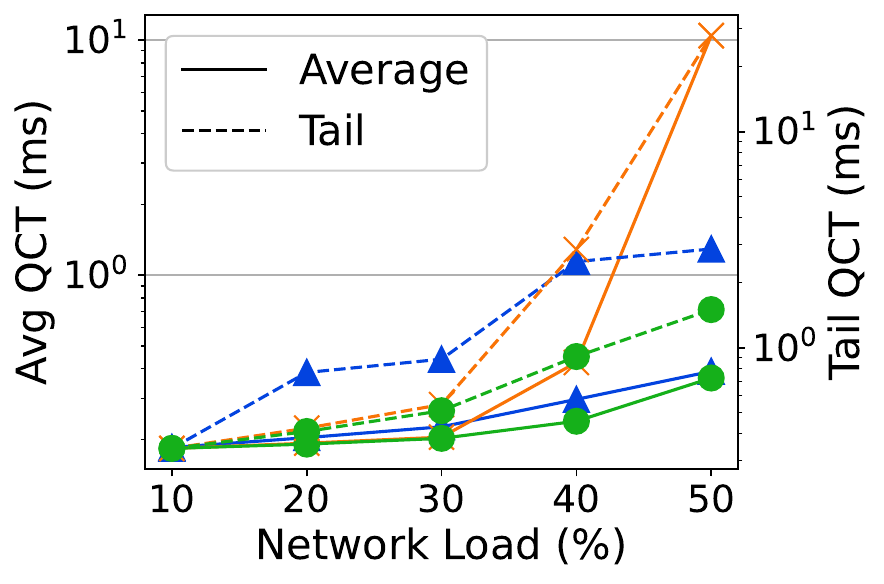}
            \caption{Incast traffic}
            \label{fig:avg_tail_qct_incast}
        \end{subfigure}
        \begin{subfigure}{0.49\linewidth}
            \centering
            \includegraphics[width=\linewidth]{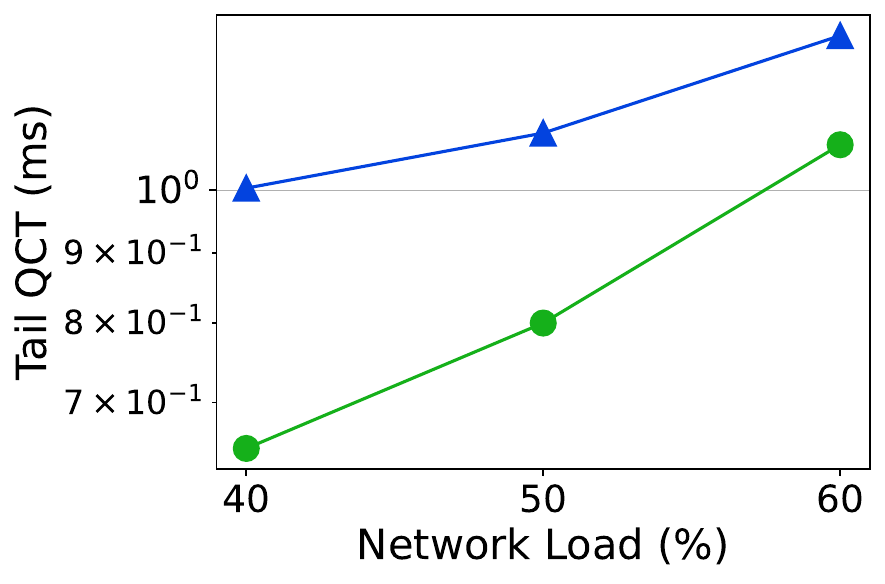}
            \caption{Incast and BG traffic}
            \label{fig:tail_qct_mix}
        \end{subfigure}
        \caption{Eunomia facilitates packet deflection in RDMA networks}
        \label{fig:incast_fig}
    \end{minipage}
    \vspace{-2mm}
\end{figure}

\noindent\textbf{Result: } Fig \ref{fig:avg_tail_qct_incast} shows that as the incast load increases, IRN fails to keep up with the excessive packet losses in the network and deteriorates performance. DIBS + IRN performs better at low network load because DIBS reduces packet drops that occur, however, as the load increases the reordering created by deflection causes the performance to degrade because IRN cannot keep up with high reordering. DIBS + Eunomia continues to perform better with up to 47\% lower average and tail query completion time (QCT) than IRN. Note that we only compare till 50\% load as it has been noticed that DIBS itself underperforms at higher loads. The goal of our experiment is to show deflection as an application of Eunomia, more sophisticated deflection techniques that work well at higher loads can be employed for their gains as Eunomia is equipped to handle high degrees of reordering.

We also experiment with mixed traffic, Fig \ref{fig:tail_qct_mix} contains tail QCT in the presence of background (BG) traffic. The incast traffic is fixed at 30\% load while the background traffic varies from 10\% to 30\%. We only include low to moderate loads for the same reason mentioned above. Eunomia allows deflection to reduce packet drops and handle incast better, resulting in a much smaller tail QCT even in the presence of background traffic.

\subsection{Eunomia enables RDMA to incorporate failure management mechanisms}
\label{sec:eval_failure}
Eunomia enables RDMA network to build resilience towards link failures.

\textbf{Setup: } We fail a percentage of network links at fixed intervals (100ms), at the end of each interval, the failed links are brought back up and a new set of links fail. Upon a link failure, the switches drain the buffers of failed link ports, recalculate the routing tables, and reroute the flows if if needed. The percentage of failed links is varied from 1-10 \%.

\noindent\textbf{Result: } In Fig \ref{fig:link_failure_eval} the x-axis represents the percentage of failed network links and the y-axis represents a slowdown in FCTs. Fig \ref{fig:link_failure_eval} shows that as the number of failures increases, the performance of ECMP deteriorates. Whereas finer-grained load balancers combined with Eunomia, hold up to a high level of failure and maintain their performance because they use multiple paths for each flow, so the number of packets a flow might lose due to failure is not high. Therefore, even when GBN is triggered to recover lost packets, the window moves forward quickly because Eunomia buffers the out-of-order packets. We omit results with Conweave because its performance falls to 0, and no flow reaches completion beyond 2\% link failure. Conweaves' ordering support relies on communication between the source and destination TOR regarding route changes, but in case of failures, a flow changes route without the destination TOR's knowledge which, consequently, fails to provide ordering support. Moreover, Conweav uses one priority queue per flow to handle reordering and faces scalability issues when more flows are reordered.

\subsection{Eunomia incurs a shallow memory footprint}

We observe the dynamic bitmap design of Eunomia delivers the same performance as an ideally sized static bitmap while incurring a significantly lower memory footprint.

\textbf{Setup: } For this experiment, finer-gained LBs we either use a static bitmap (tested with various sizes) or Eunomia (dynamic bitmap). For the first set of experiments, we vary the size of the static bitmap to evaluate performance with different sizes. In the latter set, we used the \textit{best lowest size} picked from the first set.

\begin{figure}
\begin{minipage}[b]{0.64\linewidth}
    \centering
  \begin{subfigure}{0.32\linewidth}
    \centering
    \includegraphics[width=\linewidth]{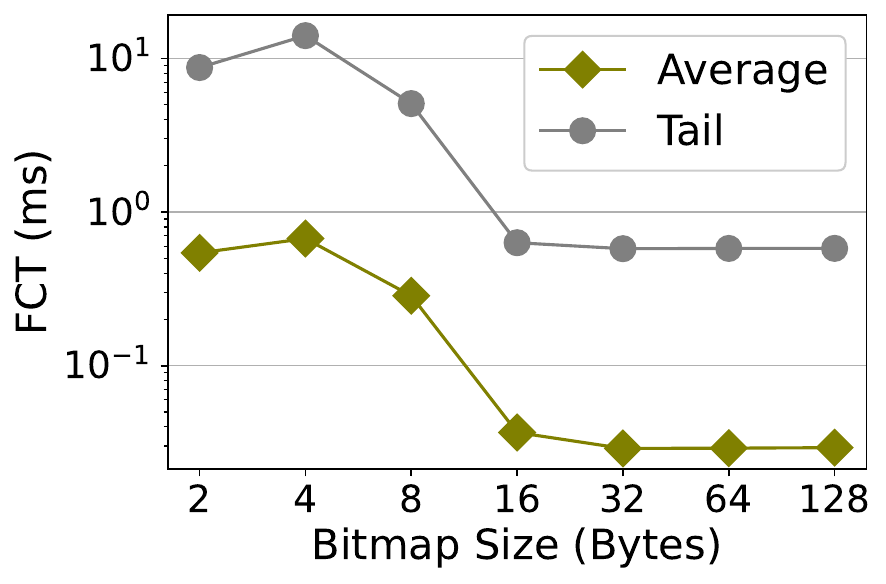}
    \caption{FCT with Static Bitmap at 80\% Load}
    \label{fig:static_bitmap_drill_80p}
  \end{subfigure}
  \hfill
  \begin{subfigure}{0.32\linewidth}
    \centering
    \includegraphics[width=\linewidth]{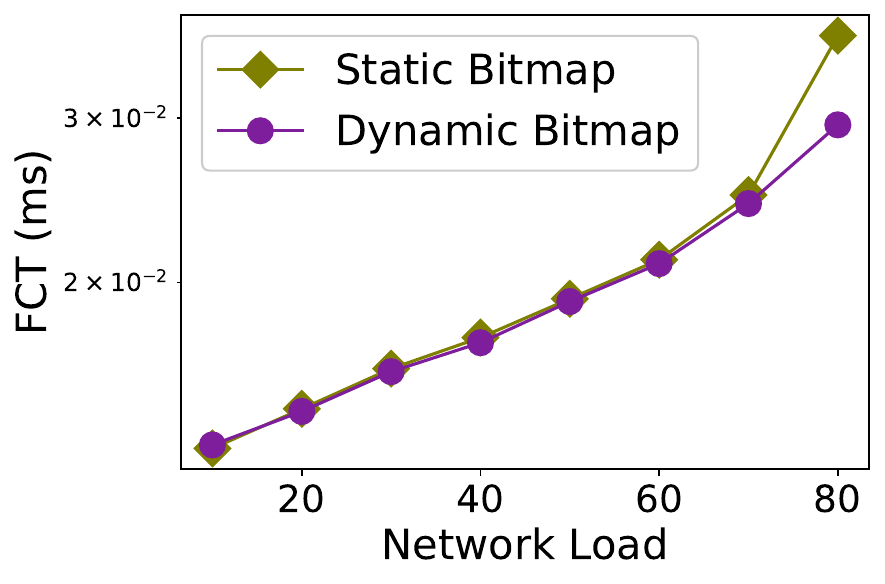}
    \caption{Avg FCT at varying load}
    \label{fig:static_vs_dynamic_fct}
  \end{subfigure}
  \hfill
  \begin{subfigure}{0.32\linewidth}
    \centering
    \includegraphics[width=\linewidth]{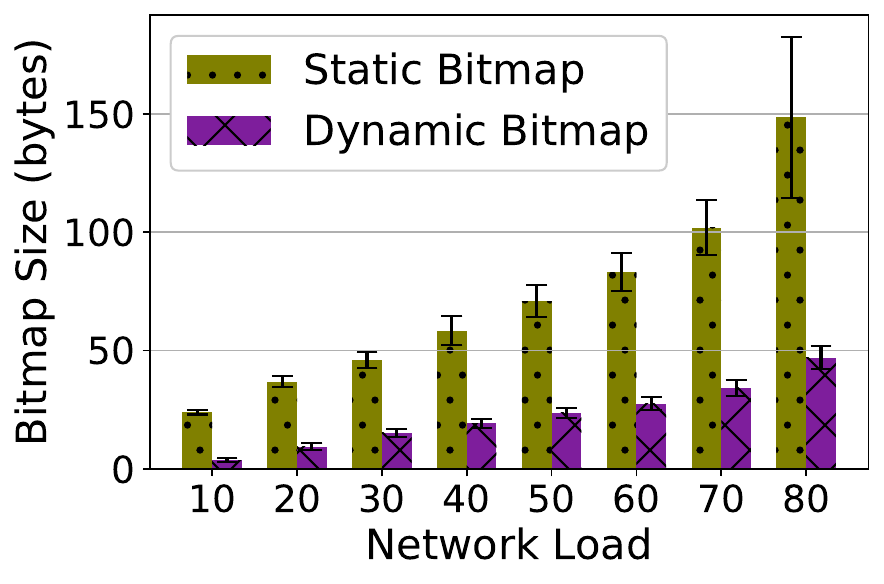}
    \caption{Memory Footprint at varying load}
    \label{fig:static_vs_dynamic_memory}
  \end{subfigure}
  \vspace{-2mm}
  \caption{Eunomia provides similar and even better performance than an appropriately sized static bitmap with a significantly lower memory footprint in Clos topology with DRILL LB under AliStorage workload.
  }
  \label{fig:static_vs_dynamic}
  \end{minipage}
  \hfill
  \begin{minipage}[b]{0.34\linewidth}
        \centering
        \begin{subfigure}[b]{\linewidth}
            \centering
            \includegraphics[width=\linewidth]{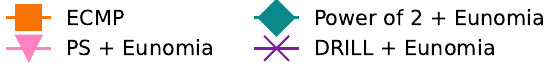}
        \end{subfigure}
        \begin{subfigure}[b]{0.65\linewidth}
            \centering
            \includegraphics[width=\linewidth]{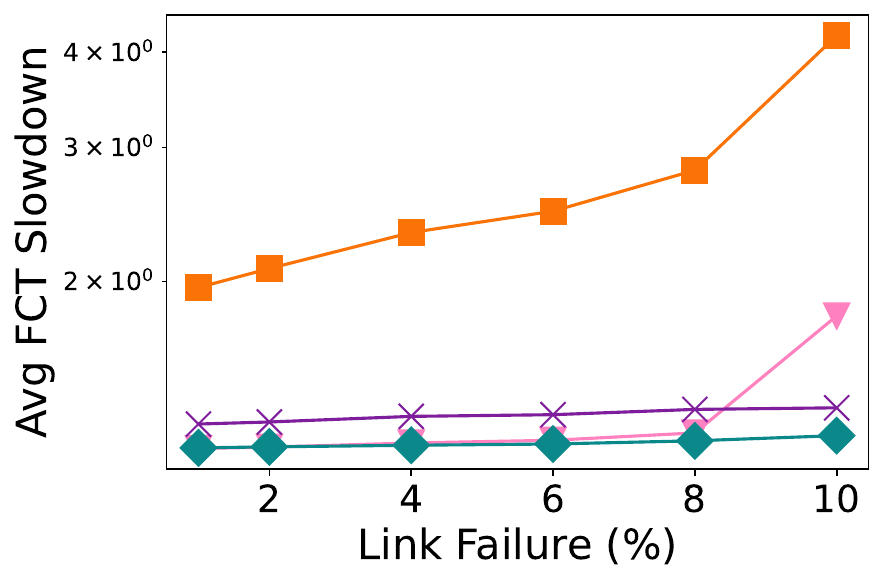}
        \end{subfigure}
        \vspace{-2mm}
        \caption{With the help of Eunomia, RDMA achieves resilience towards link failures and degrades gracefully under extreme link failures.}
        \label{fig:link_failure_eval}
    \end{minipage}
 \vspace{-2mm}
\end{figure}

\textbf{Result } First we demonstrate the impact of static bitmap size on performance, Fig \ref{fig:static_bitmap_drill_80p} shows average and tail FCT in DRILL when using different size of static bitmap. We observe high average and tail FCT with a small size of the static bitmap ($\leq8$ bytes). However, as the size increases the performance stabilizes with a bitmap of 16 bytes significantly reducing the FCTs, and it converges more with a larger size of $\geq32$ bytes. These results demonstrate that to reap the benefits of a fine-grained LB, it is important to set the size of the static bitmap correctly, which relies on a series of factors discussed in section \ref{sec:dynamic_reordering_bitmap}. For this particular setting, a size of 16 bytes seems appropriate, albeit not ideal because a 32-byte bitmap results in a lower average FCT.

We now compare the performance of the static bitmap against dynamic bitmap (Eunomia), in terms of FCT as well as the memory footprint. For static bitmap, we use a size of 16 bytes, while for Eunomia we use the same setting as used in other experiments. Fig \ref{fig:static_vs_dynamic_fct} shows that Eunomia (with its dynamic bitmap) performs similarly and even better than a corresponding static bitmap. Furthermore, the performance of Eunomia comes with a significantly lower memory footprint as compared to a static bitmap, as seen in Fig \ref{fig:static_vs_dynamic_memory}, which shows the the average total bytes taken by all bitmaps on a NIC measured at small intervals (1ms). The reason for similar performance but a significantly lower memory footprint of Eunomia is due to the efficient management of resources via dynamic bitmap, by only allocating as many bytes for bitmaps as needed for each flow, while the static bitmap is forced to allocate the same amount of bytes to all flows despite their degree of reordering.

\section{Related Work}
In this section, we summarize the prior research on providing ordering support in RDMAs, including switch-based (e.g., Conweave \cite{song2023network}) and NIC-based designs (e.g., MPRDMA \cite{lu2018multi} and IRN \cite{mittal2018revisiting}).

\textbf{Conweave \cite{song2023network}} is an LB and comes under the switch-based designs, it is implemented on the top-of-rack (TOR) switches where the source TOR is responsible for load balancing and the destination TOR for packet reordering. While it enables RDMA networks to somewhat leverage the multi-path topology, it is quite limited as it can only utilize up to two paths per flow. Furthermore, it uses priority queues at the destination TOR to track OOO packets, one queue per flow. So, the number of flows that can use multi-paths is limited by the number of priority queues at the destination TOR. Lastly, Conweave does not offer a standalone reordering handler to be used with other applications.

\textbf{MP-RDMA \cite{lu2018multi}} a multi-path congestion control for RDMA is a NIC-based design, similar to MP-TCP \cite{raiciu2011improving} for TCP/IP datacenters, which handles packet reordering at the receiver NIC using a small fixed-sized bitmap. Due to the memory constraint on the NIC, the bitmap size is limited, so MP-RDMA aims to keep packet reordering low by pruning "slower" paths. This limits the available paths for a flow and therefore MP-RDMA fails to utilize the network to its fullest. Furthermore, the performance of MPRDMA is sensitive to parameter tuning, which we observed in Fig \ref{fig:motivation_clos_avg_fct_80p} where it provides considerable gains compared to ECMP, but falls short compared to Conweave \cite{song2023network} and other finer-grained LBs. Many of the larger flows do not reach completion due to synchronization failure (which is directly affected by a particular parameter). The mid-size flows whose sizes are smaller than the initial window (another tunable parameter, 1BDP recommended) but larger than the reordering bitmap suffer as well because MPRDMA fails to control their degree of reordering.


\textbf{IRN \cite{mittal2018revisiting} } is a NIC-based design that lifts the losslessness requirement of RDMA by equipping the RDMA NICs with a simple form of selective repeat (SR) recovery mechanism that absorbs the OOO packets caused by packet loss. IRN also uses flow control to manage switch queue build-up and avoid packet loss. The existing design of IRN is configured only to handle packet loss, any packet that is received OOO is assumed to indicate a packet loss and consequently puts the sender into loss recovery. This can hinder the overall performance when there is high reordering in the network induced by some scheme (say a fine-grained LB). This hypothesis is confirmed in our experiments shown in Fig \ref{fig:motivation_clos_avg_fct_80p_irn} and in section \ref{sec:eval_fine_grain_lb} that fine-grained LBs degrade considerably in performance as opposed to ECMP and Conweave, despite IRN using selective repeat, indicating that IRN alone can not handle high levels of packet reordering.


\section{Conclusion}
In this paper, we highlighted the limitations of RDMA datacenter networks that arise due to the in-order packet delivery constraint. We proposed Eunomia, an on-NIC design to provide ordering support in RDMA datacenter networks. We implemented and tested the feasibility of our design in FPGA. We also evaluated Eunomia via large-scale simulations. Our results show that Eunomia lifts the in-order delivery constraint of RDMA networks and enables various performance-enhancing techniques in RDMA, providing significant performance gains. 
\bibliography{mybibliography} 

\begin{thebibliography}{10}

\bibitem{li2019hpcc}
Y.~Li, R.~Miao, H.~H. Liu, Y.~Zhuang, F.~Feng, L.~Tang, Z.~Cao, M.~Zhang, F.~Kelly, M.~Alizadeh, {\em et~al.}, ``Hpcc: High precision congestion control,'' in {\em Proceedings of the ACM Special Interest Group on Data Communication}, pp.~44--58, 2019.

\bibitem{guo2016rdma}
C.~Guo, H.~Wu, Z.~Deng, G.~Soni, J.~Ye, J.~Padhye, and M.~Lipshteyn, ``Rdma over commodity ethernet at scale,'' in {\em Proceedings of the 2016 ACM SIGCOMM Conference}, pp.~202--215, 2016.

\bibitem{pfc}
{IEEE}, ``{IEEE Std 802.1Qbb-2011}: Standard for local and metropolitan area networks--bridges and bridged networks--amendment 25: Priority-based flow control,'' Standard 802.1Qbb, 2011.

\bibitem{bfc}
P.~Goyal, P.~Shah, N.~K. Sharma, M.~Alizadeh, and T.~E. Anderson, ``Backpressure flow control,'' in {\em Proceedings of the 2019 Workshop on Buffer Sizing}, pp.~1--3, 2019.

\bibitem{valadarsky2016xpander}
A.~Valadarsky, G.~Shahaf, M.~Dinitz, and M.~Schapira, ``Xpander: Towards optimal-performance datacenters,'' in {\em Proceedings of the 12th International on Conference on emerging Networking EXperiments and Technologies}, pp.~205--219, 2016.

\bibitem{singla2012jellyfish}
A.~Singla, C.-Y. Hong, L.~Popa, and P.~B. Godfrey, ``Jellyfish: Networking data centers randomly,'' in {\em 9th USENIX Symposium on Networked Systems Design and Implementation (NSDI 12)}, pp.~225--238, 2012.

\bibitem{rps}
A.~Dixit, P.~Prakash, Y.~C. Hu, and R.~R. Kompella, ``On the impact of packet spraying in data center networks,'' in {\em 2013 proceedings ieee infocom}, pp.~2130--2138, IEEE, 2013.

\bibitem{ghorbani2017drill}
S.~Ghorbani, Z.~Yang, P.~B. Godfrey, Y.~Ganjali, and A.~Firoozshahian, ``Drill: Micro load balancing for low-latency data center networks,'' in {\em Proceedings of the Conference of the ACM Special Interest Group on Data Communication}, pp.~225--238, 2017.

\bibitem{presto}
K.~He, E.~Rozner, K.~Agarwal, W.~Felter, J.~Carter, and A.~Akella, ``Presto: Edge-based load balancing for fast datacenter networks,'' {\em ACM SIGCOMM Computer Communication Review}, vol.~45, no.~4, pp.~465--478, 2015.

\bibitem{homa}
B.~Montazeri, Y.~Li, M.~Alizadeh, and J.~Ousterhout, ``Homa: A receiver-driven low-latency transport protocol using network priorities,'' in {\em Proceedings of the 2018 Conference of the ACM Special Interest Group on Data Communication}, pp.~221--235, 2018.

\bibitem{ndp}
C.~Raiciu and G.~Antichi, ``Ndp: Rethinking datacenter networks and stacks two years after,'' {\em ACM SIGCOMM Computer Communication Review}, vol.~49, no.~5, pp.~112--114, 2019.

\bibitem{abdous2021burst}
S.~Abdous, E.~Sharafzadeh, and S.~Ghorbani, ``Burst-tolerant datacenter networks with vertigo,'' in {\em Proceedings of the 17th International Conference on emerging Networking EXperiments and Technologies}, pp.~1--15, 2021.

\bibitem{canary}
S.~Abdous, E.~Sharafzadeh, and S.~Ghorbani, ``Practical packet deflection in datacenters,'' {\em Proceedings of the ACM on Networking}, vol.~1, no.~CoNEXT3, pp.~1--25, 2023.

\bibitem{vanini2017let}
E.~Vanini, R.~Pan, M.~Alizadeh, P.~Taheri, and T.~Edsall, ``Let it flow: Resilient asymmetric load balancing with flowlet switching,'' in {\em 14th USENIX Symposium on Networked Systems Design and Implementation (NSDI 17)}, pp.~407--420, 2017.

\bibitem{dibs}
K.~Zarifis, R.~Miao, M.~Calder, E.~Katz-Bassett, M.~Yu, and J.~Padhye, ``Dibs: Just-in-time congestion mitigation for data centers,'' in {\em Proceedings of the Ninth European Conference on Computer Systems}, pp.~1--14, 2014.

\bibitem{alizadeh2013pfabric}
M.~Alizadeh, S.~Yang, M.~Sharif, S.~Katti, N.~McKeown, B.~Prabhakar, and S.~Shenker, ``pfabric: Minimal near-optimal datacenter transport,'' {\em ACM SIGCOMM Computer Communication Review}, vol.~43, no.~4, pp.~435--446, 2013.

\bibitem{zhu2015congestion}
Y.~Zhu, H.~Eran, D.~Firestone, C.~Guo, M.~Lipshteyn, Y.~Liron, J.~Padhye, S.~Raindel, M.~H. Yahia, and M.~Zhang, ``Congestion control for large-scale rdma deployments,'' {\em ACM SIGCOMM Computer Communication Review}, vol.~45, no.~4, pp.~523--536, 2015.

\bibitem{mittal2018revisiting}
R.~Mittal, A.~Shpiner, A.~Panda, E.~Zahavi, A.~Krishnamurthy, S.~Ratnasamy, and S.~Shenker, ``Revisiting network support for rdma,'' in {\em Proceedings of the 2018 Conference of the ACM Special Interest Group on Data Communication}, pp.~313--326, 2018.

\bibitem{alizadeh2014conga}
M.~Alizadeh, T.~Edsall, S.~Dharmapurikar, R.~Vaidyanathan, K.~Chu, A.~Fingerhut, V.~T. Lam, F.~Matus, R.~Pan, N.~Yadav, {\em et~al.}, ``Conga: Distributed congestion-aware load balancing for datacenters,'' in {\em Proceedings of the 2014 ACM conference on SIGCOMM}, pp.~503--514, 2014.

\bibitem{pabo}
X.~Shi, L.~Wang, F.~Zhang, K.~Zheng, M.~M{\"u}hlh{\"a}user, and Z.~Liu, ``Pabo: Mitigating congestion via packet bounce in data center networks,'' {\em Computer Communications}, vol.~140, pp.~1--14, 2019.

\bibitem{lu2018multi}
Y.~Lu, G.~Chen, B.~Li, K.~Tan, Y.~Xiong, P.~Cheng, J.~Zhang, E.~Chen, and T.~Moscibroda, ``$\{$Multi-Path$\}$ transport for $\{$RDMA$\}$ in datacenters,'' in {\em 15th USENIX symposium on networked systems design and implementation (NSDI 18)}, pp.~357--371, 2018.

\bibitem{song2023network}
C.~H. Song, X.~Z. Khooi, R.~Joshi, I.~Choi, J.~Li, and M.~C. Chan, ``Network load balancing with in-network reordering support for rdma,'' in {\em Proceedings of the ACM SIGCOMM 2023 Conference}, pp.~816--831, 2023.

\bibitem{180325}
V.~Liu, D.~Halperin, A.~Krishnamurthy, and T.~Anderson, ``F10: A {Fault-Tolerant} engineered network,'' in {\em 10th USENIX Symposium on Networked Systems Design and Implementation (NSDI 13)}, (Lombard, IL), pp.~399--412, USENIX Association, Apr. 2013.

\bibitem{ns3_simul}
nsnam, ``nsnam,'' 2024.
\newblock Accessed: 2024-06-08.

\bibitem{raiciu2011improving}
C.~Raiciu, S.~Barre, C.~Pluntke, A.~Greenhalgh, D.~Wischik, and M.~Handley, ``Improving datacenter performance and robustness with multipath tcp,'' {\em ACM SIGCOMM Computer Communication Review}, vol.~41, no.~4, pp.~266--277, 2011.

\bibitem{gill2011understanding}
P.~Gill, N.~Jain, and N.~Nagappan, ``Understanding network failures in data centers: measurement, analysis, and implications,'' in {\em Proceedings of the ACM SIGCOMM 2011 Conference}, pp.~350--361, 2011.

\bibitem{wang2023srnic}
Z.~Wang, L.~Luo, Q.~Ning, C.~Zeng, W.~Li, X.~Wan, P.~Xie, T.~Feng, K.~Cheng, X.~Geng, {\em et~al.}, ``$\{$SRNIC$\}$: A scalable architecture for $\{$RDMA$\}$$\{$NICs$\}$,'' in {\em 20th USENIX Symposium on Networked Systems Design and Implementation (NSDI 23)}, pp.~1--14, 2023.

\bibitem{kalia2019datacenter}
A.~Kalia, M.~Kaminsky, and D.~Andersen, ``Datacenter $\{$RPCs$\}$ can be general and fast,'' in {\em 16th USENIX Symposium on Networked Systems Design and Implementation (NSDI 19)}, pp.~1--16, 2019.

\bibitem{nvidia-connectx5}
{NVIDIA Corporation}, {\em {NVIDIA ConnectX-5 Ethernet Adapter Cards User Manual}}.
\newblock {NVIDIA Corporation}.

\bibitem{roy2015inside}
A.~Roy, H.~Zeng, J.~Bagga, G.~Porter, and A.~C. Snoeren, ``Inside the social network's (datacenter) network,'' in {\em Proceedings of the 2015 ACM Conference on Special Interest Group on Data Communication}, pp.~123--137, 2015.

\end{thebibliography}
\bibliographystyle{ieeetr}

\appendix
\newpage
\section{Appendix}
\subsection{WRITE Verb with Eunomia}
\begin{wrapfigure}{r}{0.25\textwidth} 
    \centering
    \vspace{-4mm} 
    \begin{subfigure}[b]{\linewidth}
            \centering
            \includegraphics[width=\linewidth]{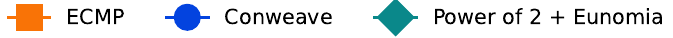}
        \end{subfigure}
        \begin{subfigure}{\linewidth}
            \centering
            \includegraphics[width=\linewidth]{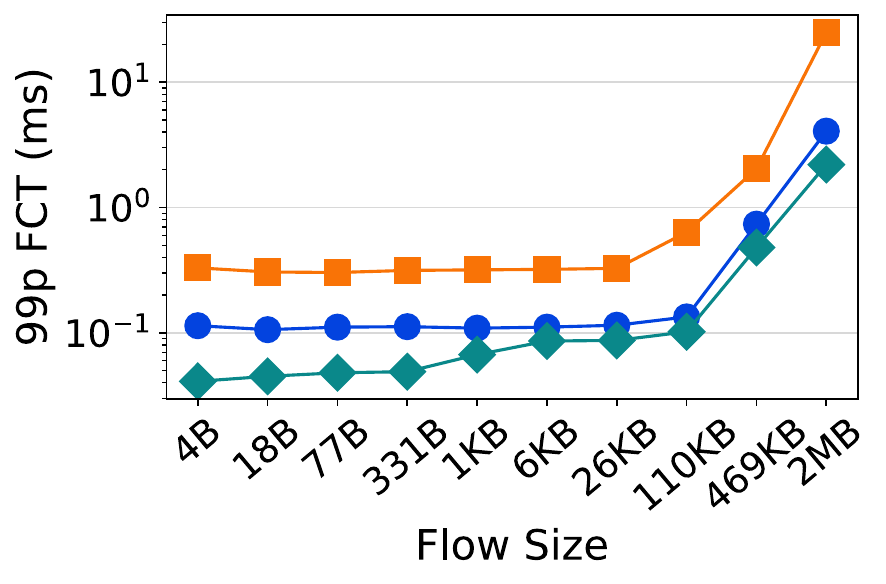}
            \caption{Tail FCT at 80\% Load}
            \end{subfigure}

        \caption{Eunomia allows RDMA to gain better performance for WRITE verbs despite the additional delay}
        \label{fig:clos_fct_write}
    \vspace{-3mm}
\end{wrapfigure}
Eunomia ensures in-order delivery to the application by delaying the completion notification (CN) until everything is received in order. However, WRITE verbs in RDMA are one-sided operations and do not use CN. For such a case, Eunomia's send side agent delays sending the last packet of a connection until everything up until that packet has been received in order, which adds an RTT worth of delay to the flow completion time. We run a set of experiments with the same setup as \S\ref{sec:eval_fine_grain_lb} except for a slight difference, all flows in the traffic are WRITE verb connections and therefore Eunomia delays sending their last packet to be received in order. Fig \ref{fig:clos_fct_write} shows that at high network load, Eunomia allows for better performance than ECMP even with the additional delay of 1 RTT. However, despite allowing RDMA to perform better than the current schemes, it is not desirable to add an extra RTT, especially for short-lived connections. We propose another solution to this problem, which is to handle the last OOO packet at the receiver NIC. The receiver side agent holds off the last packet of a WRITE connection in the host memory if it is received OOO and only places it in the application memory when everything leading up to that packet is received in order. This takes up extra PCIe bandwidth for the last packet but would incur a smaller delay than 1 RTT. We leave this addition to Eunomia for future work.

\end{document}